\documentclass[twocolumn]{aastex631}

\tablewidth{8in}

\usepackage{amsmath}
\usepackage{aasmacros} 
\graphicspath{{Figures/}}
\usepackage{duckuments} 
\usepackage{soul} 

\newcommand{\vanderbilt}{Department of Physics and Astronomy,
Vanderbilt University, 6301 Stevenson Center, Nashville, TN 37235}

\newcommand{\utk}{Department of Physics and Astronomy,
University of Tennessee, Knoxville, Nielsen Physics Building, 401,
1408 Circle Drive, Knoxville, TN 37996}

\newcommand{\ornl}{Computer Science and Mathematics Division,
Oak Ridge National Laboratory, Oak Ridge, TN 37831}

\newcommand{\ncsu}{Department of Physics,
North Carolina State University, 2401 Stinson Dr., Raleigh, NC 27607}

\newcommand{\fisk}{Department of Life and Physical Sciences,
Fisk University, 1000 17th Ave N, Nasvhille, TN 37208}

\hypersetup{
  bookmarksnumbered = true,
  bookmarksopen     = false,
  pdfborder         = 0 0 0,
  pdffitwindow      = true,
  pdfnewwindow      = true,
  colorlinks        = true,
  linkcolor         = blue,
  citecolor         = magenta,
  filecolor         = magenta,
  urlcolor          = cyan
}

\accepted{January 19, 2024}
\published{March 14, 2024}
\submitjournal{ApJ}

\shorttitle{GR SASI}
\shortauthors{Dunham et al.}

\newcommand{\PeriodRatioGRoverNRxiA}{1.19}
\newcommand{\PeriodRatioGRoverNRxiB}{1.15}
\newcommand{\PeriodRatioGRoverNRxiC}{1.22}
\newcommand{\PeriodRatioGRoverNRxiD}{1.29}

\newcommand{\GrowthRateRatioGRoverNRxiD}{0.47}
\newcommand{\RelDiffAAxiA}{0.29}
\newcommand{\RelDiffAAxiB}{0.14}
\newcommand{\RelDiffAAxiC}{0.05}
\newcommand{\RelDiffAAxiD}{0.17}
\newcommand{\RelDiffACxiA}{0.21}
\newcommand{\RelDiffACxiB}{0.22}
\newcommand{\RelDiffACxiC}{0.10}
\newcommand{\RelDiffACxiD}{0.12}

\newcommand{\p}{\partial}
\newcommand{\pd}[2]{\frac{\partial#1}{\partial#2}}

\newcommand{\T}{^{\top}}

\newcommand{\wt}{\widetilde}
\newcommand{\etac}{\eta_{\mathrm{c}}}
\newcommand{\rpns}{R_{\textsc{pns}}}
\newcommand{\rsh}{R_{\mathrm{sh}}}
\newcommand{\rac}{R_{\mathrm{ac}}}
\newcommand{\rsc}{R_{\mathrm{Sc}}}
\newcommand{\gammabar}{\overline{\gamma}}
\newcommand{\myeqref}[1]{Eq.~(\ref{#1})}
\newcommand{\thornado}{\texttt{thornado}}
\newcommand{\amrex}{\texttt{AMReX}}
\newcommand{\eos}{EoS}
\newcommand{\cfc}{CFC}

\newcommand{\msun}{M_{\odot}}
\newcommand{\mdot}{\dot{M}}
\newcommand{\gr}{\textsc{gr}}
\newcommand{\nr}{\textsc{nr}}
\newcommand{\tauad}{\tau_{\mathrm{ad}}}
\newcommand{\tauac}{\tau_{\mathrm{ac}}}
\newcommand{\tac}{T_{\mathrm{ac}}}
\newcommand{\taa}{T_{\mathrm{aa}}}
\newcommand{\mpns}{M_{\textsc{pns}}}
\newcommand{\bs}[1]{\boldsymbol{#1}}
\newcommand{\mc}{\mathcal}
\newcommand{\bb}{\mathbb}

\newcommand{\figref}[1]{Figure~\ref{#1}}
\newcommand{\tabref}[1]{Table~\ref{#1}}
\newcommand{\secref}[1]{Section~\ref{#1}}
\renewcommand{\th}{-th}
\newcommand{\nModels}{7}
\newcommand{\scipy}{\texttt{scipy}}
\newcommand{\curvefit}{\texttt{curve\_fit}}

\newcommand{\erg}{\mathrm{erg}}
\newcommand{\g}{\mathrm{g}}
\newcommand{\cm}{\mathrm{cm}}
\newcommand{\km}{\mathrm{km}}
\newcommand{\ms}{\mathrm{ms}}
\newcommand{\s}{\mathrm{s}}

\newcommand{\rad}{\mathrm{rad}}

\begin{document}


\title{%
A Parametric Study of the SASI Comparing General Relativistic and Nonrelativistic Treatments\footnote{Notice:  This manuscript has been authored by UT-Battelle, LLC, under contract DE-AC05-00OR22725 with the US Department of Energy (DOE). The US government retains and the publisher, by accepting the article for publication, acknowledges that the US government retains a nonexclusive, paid-up, irrevocable, worldwide license to publish or reproduce the published form of this manuscript, or allow others to do so, for US government purposes. DOE will provide public access to these results of federally sponsored research in accordance with the DOE Public Access Plan (http://energy.gov/downloads/doe-public-access-plan).}}

\correspondingauthor{Samuel J. Dunham}
\email{samuel.j.dunham@vanderbilt.edu}

\author[0000-0003-4008-6438]{Samuel J. Dunham}
\affiliation{\vanderbilt}
\affiliation{\utk}

\author[0000-0003-1251-9507]{Eirik Endeve}
\affiliation{\ornl}
\affiliation{\utk}

\author[0000-0001-9816-9741]{Anthony Mezzacappa}
\affiliation{\utk}

\author[0000-0001-9691-6803]{John M. Blondin}
\affiliation{\ncsu}

\author[0000-0002-7712-5835]{Jesse Buffaloe}
\affiliation{\utk}

\author[0000-0003-2227-1322]{Kelly Holley-Bockelmann}
\affiliation{\vanderbilt}
\affiliation{\fisk}

\begin{abstract}
We present numerical results from a parameter study of the standing accretion shock instability (SASI), investigating the impact of general relativity (GR) on the dynamics. Using GR hydrodynamics with GR gravity, and nonrelativistic (NR) hydrodynamics with Newtonian gravity, in an idealized model setting, we vary the initial radius of the shock and, by varying its mass and radius in concert, the proto-neutron star (PNS) compactness. We investigate four compactnesses expected in a post-bounce core-collapse supernova (CCSN). We find that GR leads to a longer SASI oscillation period, with ratios between the GR and NR cases as large as \PeriodRatioGRoverNRxiD{} for the highest-compactness suite. We also find that GR leads to a slower SASI growth rate, with ratios between the GR and NR cases as low as \GrowthRateRatioGRoverNRxiD{} for the highest-compactness suite. We discuss implications of our results for CCSN simulations.
\end{abstract}



\section{Introduction}

Since the discovery of the standing accretion shock instability
\citep[SASI;][]{bmd2003}, which many two- and three-dimensional simulations
performed to date demonstrate becomes manifest during a
core-collapse supernova (CCSN) in the post-shock accretion flow
onto the proto-neutron star (PNS),
groups have made efforts to understand its physical origin
and its effects on the supernova itself.
The SASI is characterized in two dimensions (2D) by a large-scale
``sloshing" of the shocked fluid,
and in three dimensions (3D) by additional spiral modes \citep{bm2007}.
It is now generally accepted that turbulent neutrino-driven convection
plays a major role in re-energizing the stalled shock
\citep[e.g., see][]{%
bdm2012,
hmw2013,
mdb2013,
co2015,
mjm2015,
lbh2015,
aor2015,
rco2015,
mjm2015,
mjb2015,
roh2016,
mmh2017,
sjm2018,
rao2018,
oc2018b,
vbr2019,
mth2019,
brv2019,
ytk2019,
pm2020,
sjk2020,
mv2020,
vcb2022,
ntk2022,
mat2022%
}.
The same simulations that led to the above conclusion also generally exhibit
the SASI, with outcomes ranging from convection-dominated flows to
SASI-dominated flows, and flows where neither dominates.
Strong SASI activity and, in some cases, SASI-aided explosions have been
reported in, for example, the three-dimensional simulations of
\cite{sjm2018}, \cite{oc2018b}, and \cite{mat2022}.
A more precise determination of the relative role played by these two
instabilities in the explosion mechanism, on a case-by-case basis
\citep[i.e., for different progenitor characteristics; e.g., see][]{%
sjf2008,
hmm2012,
hmw2013,
co2014,
fmf2014,
mjb2015,
aor2015,
f2015%
}, will require advances in current three-dimensional models to include full
general relativity,
rotation, magnetic fields, and the requisite neutrino interaction physics
with realistic spectral neutrino kinetics, all at high spatial resolution.
It is also important to note that, while
convection-dominated and SASI-dominated scenarios may lie at the extremes of
what is possible, it is not necessary for one or the other instability to be
dominant to play an important role -- specifically, for the more complex cases
where neither dominates,
it would be very difficult
to determine precisely the relative
contribution from these two instabilities.

Several studies have concluded that the SASI is an
{\sl  advective-acoustic} instability, in which
vortical waves generated at the shock advect to the surface of the PNS,
which in turn generate acoustic waves that propagate back to the
shock and further perturb it
\citep{fsj2006,fgs2007,yy2007,l2007,l2008,f2009,gf2012}.
This perturbation generates more, stronger, vortical waves,
which advect to
the PNS surface, thus creating a feedback loop that drives the instability.
An alternative explanation for the SASI is the
{\sl purely acoustic} mechanism, in which
acoustic perturbations just below the shock travel around
the post-shock region and constructively interfere with each other,
generating stronger acoustic perturbations
and thereby feeding the instability
\citep{bm2006}.
A recent study \citep{wft2023} suggests that the acoustic mechanism may play
a particularly important role in the SASI when rotation is included,
implying that the origins of the SASI excitation may depend on conditions
between the shock and the PNS.

Other numerical studies focus on particular aspects of this instability,
such as the hydrodynamics of the SASI
\citep{oky2006,sff2009,iny2014},
spiral modes
\citep{bs2007,iko2008,f2010},
the spin-up of the possible remnant pulsar
\citep{bm2007},
the effect of nuclear dissociation
\citep{ft2009},
saturation of the instability
\citep{gsf2010},
the generation and amplification of magnetic fields
\citep{ecb2010,ecb2012},
the relative importance of the SASI and convection in CCSNe
\citep{cb2015},
the generation of, and impact on, gravitational waves by the SASI
\citep{%
koy2007,
kio2009,
kkt2016,
a2017,
kkt2017,
amm2017,
oc2018b,
hkk2018,
amj2019,
mml2020,
mml2023,
dad2023},
and the effects of rotation
\citep{yy2005,yf2008,wft2023,bfg2023}.
Some of these studies included sophisticated microphysics, such as
realistic equations of state (\eos s) and neutrino transport;
however, with the exception of \citet{kkt2017},
none of these studies solved the general relativistic hydrodynamics (GRHD)
equations, instead solving their nonrelativistic (NRHD) counterparts,
some with an approximate relativistic gravitational potential.
It has been demonstrated that GR effects are crucial to include in CCSN
simulations \citep{bdm2001,mjm2012,lmm2012,oc2018a}, yet the SASI itself
has not been fully investigated in the GR regime.
A recent paper \citep{kc2022} does analyze steady state accretion
through a stationary shock onto compact objects in a
Schwarzchild geometry and compares with Newtonian solutions, and posits
that GR may have a non-negligible impact on the SASI.
They find that, for conditions expected in exploding CCSNe,
the freefall speed is of order $v\sim0.2\,c$
(with $c$ the speed of light), and
the differences between the GR and NR solutions are of order 10\%.
For conditions expected in failed CCSNe (i.e., supernovae where the
shock is not revived, in which case the freefall speed can be
$v\sim0.66\,c$), the differences can be larger.

The timescales that likely influence the SASI depend on signal speeds
associated with advective and acoustic modes in the region
between the shock and the PNS surface
\citep{bm2006,fgs2007,m2020}.
Motivated in part by \citet{dem2020} and \citet{kc2022},
we expect SASI simulations to behave differently depending
on whether or not the treatment of the hydrodynamics and gravity
are general relativistic.
Indeed, \citet{dem2020} presented GR steady-state solutions and
presented preliminary results from GR SASI simulations,
but did not compare results from NR and GR simulations.
Specifically, we expect both advective and acoustic modes to be
influenced by the different post-shock structure in the GR case
relative to the NR case.

This leads to our main science question:
\emph{How does a general relativistic treatment of hydrodynamics and gravity
affect the oscillation period and growth rate
of the SASI?}
To begin to address this, we
present the first comparison of the SASI in both a nonrelativistic
and a general relativistic framework, using an idealized model
with four compactnesses, meant to span the range of conditions
expected in CCSN simulations.
We focus our attention on the linear regime and characterize the SASI by its
growth rate and oscillation period, as was done in \citet{bm2006}.
To capture both the linear regime of the SASI and its transition to the
nonlinear regime,
we perform our assessment via seven axisymmetric numerical
simulations using GRHD and GR gravity, with the PNS represented by a point mass
and gravity encoded in a Schwarzchild spacetime metric.
To better assess the impact of GR, we also perform simulations
using the same parameter sets but
with NRHD and Newtonian gravity, again with the PNS represented by a point
mass, but in this case gravity is encoded in the Newtonian potential.
We will often use ``NR" to refer to the case of Newtonian gravity
and NRHD.

We use a system of units in which $c=G=1$ and also make use of the Einstein
summation convention, with Greek indices running from 0 to 3 and Latin indices
running from 1 to 3.

\section{Physical Model}

\subsection{Relativistic Gravity: Conformally-Flat Condition}

We use the 3+1 decomposition of spacetime
\citep[see, e.g.,][for details]{bfi1997,g2012,rz2013},
which, in the coordinate system
$x=\left(t,x^{i}\right)$,
introduces four degrees of freedom:
the lapse function, $\alpha\left(x\right)$,
and the three components of the shift vector,
$\beta^{i}\left(x\right)$.
We further specialize to the conformally-flat condition
\citep[\cfc{},][]{wmm1996},
effectively neglecting the impact of gravitational waves on
the dynamics.
This is a valid approximation when the CCSN progenitor is nonrotating
\citep{dnf2005}, as is the case for our simulations.
The \cfc{} forces the components of the
spatial three-metric, $\gamma_{ij}\left(x\right)$,
to take the form
\begin{equation}
  \gamma_{ij}
  =\psi^{4}\,\gammabar_{ij},
  \label{eq.gammaij}
\end{equation}
where $\psi\left(x\right)$ is the conformal factor
and the $\gammabar_{ij}$ are the components of a time-independent,
flat-space metric.
We choose an isotropic spherical-polar coordinate system,
as it is appropriate to our problem and is consistent with the \cfc;
the flat-space metric is
\begin{equation}
  \gammabar_{ij}\left(r,\theta\right)
  :=\mathrm{diag}\left(1,r^{2},r^{2}\,\sin^{2}\theta\right),
  \label{eq.gammabarij}
\end{equation}
and the lapse function, conformal factor, and shift vector
take the form given in \citet{bs2010},
\begin{alignat}{2}
  \alpha   \left(t,r,\theta,\varphi\right)&=\alpha   \left(r\right)
    &&:=\frac{1-\rsc/r}{1+\rsc/r},\\
  \psi     \left(t,r,\theta,\varphi\right)&=\psi     \left(r\right)
    &&:=1+\rsc/r,\label{eq.psi}\\
  \beta^{i}\left(t,r,\theta,\varphi\right)&=\beta^{i}\left(r\right)
    &&:=0,
\end{alignat}
where $r>\rsc$ is the isotropic radial coordinate measured from the
origin and $\rsc:=M/2$ is the Schwarzchild radius in
isotropic coordinates for an object of mass $M$.
The line element under a 3+1 decomposition in
isotropic coordinates takes the form
\begin{equation}
  ds^{2}=-\alpha^{2}\,dt^{2}+\gamma_{ij}\,dx^{i}\,dx^{j}.
  \label{eq.ds2}
\end{equation}
We note here that the {\sl proper} radius, $\mc{R}\left(r\right)$,
corresponding to the {\sl coordinate} radius, $r>\rsc$, is defined by
\begin{align}
  \mc{R}\left(r\right)
  :=&\int_{\rsc}^{r}\sqrt{\gamma_{11}}\,
  dr'
  =r-\rsc^{2}/r\notag\\
  &+2\,\rsc\,\log\left(r/\rsc\right)\geq r,
  \label{eq.properDistance}
\end{align}
where we used Eqs. (\ref{eq.gammaij}-\ref{eq.gammabarij})
with $\psi$ given by \myeqref{eq.psi}.
Under the \cfc{}, the square root of the determinant of the spatial
three-metric is
\begin{equation}
    \sqrt{\gamma}=\psi^{6}\,\sqrt{\gammabar}=\psi^{6}\,r^{2}\,\sin\theta.
\end{equation}

Our choice of isotropic coordinates is consistent with our implementation of general
relativistic hydrodynamics in the CFC approximation.
Our comparison of the Newtonian and general relativistic results is conducted
{\em within this gauge}.
(In practice, the hydrodynamics equations in the Newtonian limit in the isotropic gauge
are identical {\em in form} to the standard Newtonian hydrodynamics equations,
and we were able to run our NR simulations with our Newtonian hydrodynamics code.)

\subsection{Relativistic Hydrodynamics}

We solve the relativistic hydrodynamics equations
of a perfect fluid (i.e., no viscosity or heat transfer)
in the Valencia formulation
\citep{bfi1997,rz2013}, in which they take the form of a
system of hyperbolic conservation laws with sources. Under our assumption of
a stationary spacetime, the equations can be written as
\begin{equation}\label{eq.cLaws}
  \p_{t}\,\bs{U}
  +\frac{1}{\sqrt{\gamma}}\,
  \p_{i}\left(\alpha\,\sqrt{\gamma}\,\bs{F}^{i}\right)=\alpha\,\bs{S},
\end{equation}
where $\bs{U}:=\bs{U}\left(t,r,\theta,\varphi\right)$
is the vector of evolved fluid fields,
\begin{equation}
  \bs{U}
  =\begin{pmatrix}
  D\\
  S_{j}\\
  \tau
  \end{pmatrix}
  =\begin{pmatrix}
  \rho\,W\\
  \rho\,h\,W^{2}\,v_{j}\\
  \rho\,h\,W^{2}-p-\rho\,W
  \end{pmatrix},
\end{equation}
$\bs{F}^{i}:=\bs{F}^{i}\left(\bs{U}\right)$ is the vector of fluxes of those
fields in the $i$\th{} spatial dimension,
\begin{equation}
  \bs{F}^{i}
  =\begin{pmatrix}
  D\,v^{i}\\
  P^{i}_{~j}\\
  S^{i}-D\,v^{i}
  \end{pmatrix},
\end{equation}
and $\bs{S}:=\bs{S}\left(\bs{U}\right)$ is the vector of sources,
\begin{equation}
  \bs{S}
  =\begin{pmatrix}
  0\\
  \frac{1}{2}\,P^{ik}\,\p_{j}\,\gamma_{ik}
  -\alpha^{-1}\left(\tau+D\right)\,\p_{j}\,\alpha\\
  -\alpha^{-1}S^{j}\,\p_{j}\,\alpha
  \end{pmatrix},
\end{equation}
where $D$ is the conserved rest-mass density,
$S_{j}$ is the component
of the Eulerian momentum density in the $j$\th{} spatial dimension,
and $\tau:=E-D$, with $E$ the Eulerian energy density.
The component of the fluid three-velocity in the $j$\th{} spatial dimension
is denoted by $v^{j}$, and
$W:=\left(1-v^{i}v_{i}\right)^{-1/2}$ is the Lorentz factor
of the fluid, both as measured by an Eulerian observer.
The relativistic specific enthalpy as measured by an observer
comoving with the fluid; i.e., a comoving observer, is
$h:=1+\left(e+p\right)/\rho$, where
$\rho$ is the baryon mass density, $e$ is the internal energy density,
and $p$ is the thermal pressure, all measured by a comoving observer.
Finally, $P^{ij}:=\rho\,h\,W^{2}\,v^{i}\,v^{j}+p\,\gamma^{ij}$,
with $\gamma^{ij}$ the inverse of $\gamma_{ij}$; i.e.,
$\gamma^{ik}\gamma_{kj}=\delta^{i}_{~j}$.
See \citet{rz2013} for more details.

We close the hydrodynamics equations
with an ideal \eos,
\begin{equation}
  p\left(e\right)=\left(\Gamma-1\right)e,
  \label{eq.idealEOS}
\end{equation}
where $\Gamma\in\left(1,2\right]$ is the ratio of specific heats.
For this study, we set $\Gamma=4/3$.
We further assume the \eos{} is that of a polytrope; i.e.,
\begin{equation}
  p\left(\rho\right)=K\,\rho^{\Gamma},
  \label{eq.polytrope}
\end{equation}
where $K>0$ is the polytropic constant,
whose logarithm can be considered a proxy for the entropy, $S$;
i.e., $S\propto\log\left(p/\rho^{\Gamma}\right)$.
The constant $K$ takes different values on either side of a
shock, in accordance with physically admissible solutions.
\myeqref{eq.polytrope} is consistent with \myeqref{eq.idealEOS}
through the first law of thermodynamics for an isentropic fluid.

\subsection{Nonrelativistic Hydrodynamics}

Under the 3+1 formalism of GR and the CFC,
the effect of gravity is encoded in the metric via the lapse function,
the conformal factor, and the shift vector, whereas with
Newtonian gravity,
the metric is that of flat space and the effect of gravity
is encoded in the Newtonian gravitational potential, $\Phi$,
\begin{equation}
  \Phi\left(r\right):=-\frac{M}{r}.
\end{equation}
Of course, the NRHD equations can be recovered from the GRHD equations
by taking appropriate limits; i.e.,
$v^{i}v_{i}\ll1$, $p,e\ll\rho$,
and $\Phi\ll1$, and setting $\alpha\approx\alpha_{\nr}:=1+\Phi$
and $\psi=1-\Phi/2$.

In the case of Newtonian gravity and NRHD, we solve\begin{equation}
  \p_{t}\,\bs{U}
  +\frac{1}{\sqrt{\gammabar}}\,
  \p_{i}\left(\sqrt{\gammabar}\,\bs{F}^{i}\right)=\bs{S},
  \label{eq.cLawsNR}
\end{equation}
where
\begin{equation}
  \bs{U}
  =\begin{pmatrix}
  D\\
  S_{j}\\
  E
  \end{pmatrix}
  =\begin{pmatrix}
  \rho\\
  \rho\,v_{j}\\
  e+\frac{1}{2}\,\rho\,v^{i}\,v_{i}
  \end{pmatrix},
\end{equation}
\begin{equation}
  \bs{F}^{i}
  =\begin{pmatrix}
  \rho\,v^{i}\\
  P^{i}_{~j}\\
  \left(E+p\right)v^{i}
  \end{pmatrix},
\end{equation}
and
\begin{equation}
  \bs{S}
  =\begin{pmatrix}
  0\\
  \frac{1}{2}\,P^{ik}\,\p_{j}\,\gammabar_{ik}
  -\rho\,\p_{j}\,\Phi\\
  -S^{j}\,\p_{j}\,\Phi
  \end{pmatrix},
\end{equation}
where
$P^{ij}:=\rho\,v^{i}\,v^{j}+p\,\gammabar^{ij}$ and we assume $\Phi$
is due only to the point source PNS,
\begin{equation}
  \Phi\left(r\right):=\frac{-\mpns}{r}.
\end{equation}

\section{Steady-State Accretion Shocks}
\label{ss.ssas}

We take initial conditions for our simulations from steady-state
solutions to \myeqref{eq.cLaws} (GR) and \myeqref{eq.cLawsNR} (NR).
To determine the steady-state solutions,
we assume the fluid distribution is spherically symmetric and time-independent.
Following \citet{bmd2003},
we consider a stationary accretion shock located at $r=\rsh$
with a PNS mass $\mpns$, PNS radius $\rpns$,
and a constant mass accretion rate
$\mdot>0$.
We assume a polytropic constant ahead of the shock, $K=2\times10^{14}\,
\left[\left(\erg\,\cm^{-3}\right)/\left(\g\,\cm^{-3}\right)^{4/3}\right]$,
chosen so that the pre-shock flow is highly supersonic
(all of our models have a pre-shock Mach number greater than 15).
Given that our steady-state solutions have constant
entropy between the PNS surface and the shock, they are convectively stable.
This enables us to isolate the SASI and study its development.

\subsection{Relativistic Steady-State Solutions}

Focusing on the equation for $D$, we find
(temporarily defining $v\equiv v^{r}$),
\begin{equation}
  \alpha\,\psi^{6}\,W \times 4\pi\,r^{2}\rho\,v=-\dot{M}.
  \label{IC.eq1}
\end{equation}
Manipulation of the equations for $D$ and $\tau$
in \myeqref{eq.cLaws} yields the relativistic Bernoulli equation,
\begin{equation}
  \alpha\,h\,W=\mc{B},
  \label{IC.eq2}
\end{equation}
where $\mc{B}>0$ is the relativistic Bernoulli constant.
At spatial infinity, the fluid is assumed to be at rest and the spacetime
curvature negligible, so $\alpha_{\infty}=W_{\infty}=1$.
Further, at spatial infinity, we assume the fluid to be cold; i.e.,
$\left(e+p\right)/\rho\ll1$,
so that $h_{\infty}=1$ and $\mc{B}_{\infty}=1$.
Since $\mc{B}$ is a constant, $\mc{B}=1$ everywhere.

Given $K$, Eqs. (\ref{eq.polytrope}), (\ref{IC.eq1}),
and (\ref{IC.eq2}) (with $\mc{B}=1$)
form a system of three equations in the three unknowns,
$\rho$, $v$, and $p$.
From initial guesses $v_{0}=-\sqrt{2\,\mpns/r}$,
$\rho_{0}=-\dot{M}/\left(4\pi\,r^{2}\,v_{0}\right)$, and
$p_{0}=K\,\rho_{0}^{\Gamma}$, the first two of which are obtained
from the Newtonian approximation at a distance $r>\rsh$ for
highly supersonic flow,
we define dimensionless variables
$u_{1}:=\rho/\rho_{0}$, $u_{2}:=v/v_{0}$, and $u_{3}:=p/p_{0}$.
These are substituted into the system of equations, which are then solved
with a Newton--Raphson algorithm to determine the state of the fluid everywhere
ahead of the shock.
To join the pre- and post-shock states of the fluid at $r=\rsh$,
we apply the relativistic Rankine--Hugoniot jump conditions
\citep[i.e., the Taub jump conditions,][]{t1948}
to obtain $\rho$, $v$, and $p$ just below the shock.
Once the state of the fluid just below the shock is found,
the polytropic constant for the post-shock fluid is computed
with \myeqref{eq.polytrope}
and the same system of equations is solved for
the state of the fluid everywhere below the shock.

\subsection{Nonrelativistic Steady-State Solutions}

The steady-state solution method for the nonrelativistic case
(taken from \citet{bmd2003}) follows a similar procedure
as the relativistic case,
except we begin from the NR equations for mass density and energy density,
\begin{align}
  \p_{t}\,\rho+\frac{1}{\sqrt{\gammabar}}\,
  \p_{i}\left(\sqrt{\gammabar}\,\rho\,v^{i}\right)
  &=0,\\
  \p_{t}\,E+\frac{1}{\sqrt{\gammabar}}\,
  \p_{i}\left[\sqrt{\gammabar}\left(E+p\right)v^{i}\right]
  &=-\rho\,v^{i}\,\p_{i}\,\Phi,
\end{align}
where $E:=e+\frac{1}{2}\,\rho\,\gammabar_{ij}\,v^{i}\,v^{j}$.
From these, and making the same assumptions as in the relativistic case,
we arrive at a system of two equations for the three unknowns,
$\rho$, $v\equiv v^{r}$, and $p$,
\begin{align}
  4\pi\,r^{2}\,\rho\,v&=-\dot{M},\label{eq.MassConservationNR}\\
  \frac{1}{2}\,v^{2}+h_{\nr}+\Phi&=\mc{B}_{\nr},
  \label{eq.BernoulliNR}
\end{align}
with $h_{\nr}=h-1=\left(e+p\right)/\rho$
the nonrelativistic specific enthalpy and $\mc{B}_{\nr}$
the nonrelativistic Bernoulli constant.
Following \citet{bmd2003}, we set $\mc{B}_{\nr}=0$.
As in the GR case, we close this system with \myeqref{eq.polytrope}.

In the nonrelativistic limit,
$\alpha\approx1+\Phi$ and $W\approx1+\frac{1}{2}\,v^{i}\,v_{i}$;
substituting these into \myeqref{IC.eq2} yields, to leading order,
$1+\mc{B}_{\nr}$,
in agreement with \myeqref{eq.BernoulliNR}.

\subsection{Comparison of NR and GR Steady-State Solutions}

\begin{figure*}[htb!]
  \centering
  \includegraphics[width=0.8\textwidth]%
  {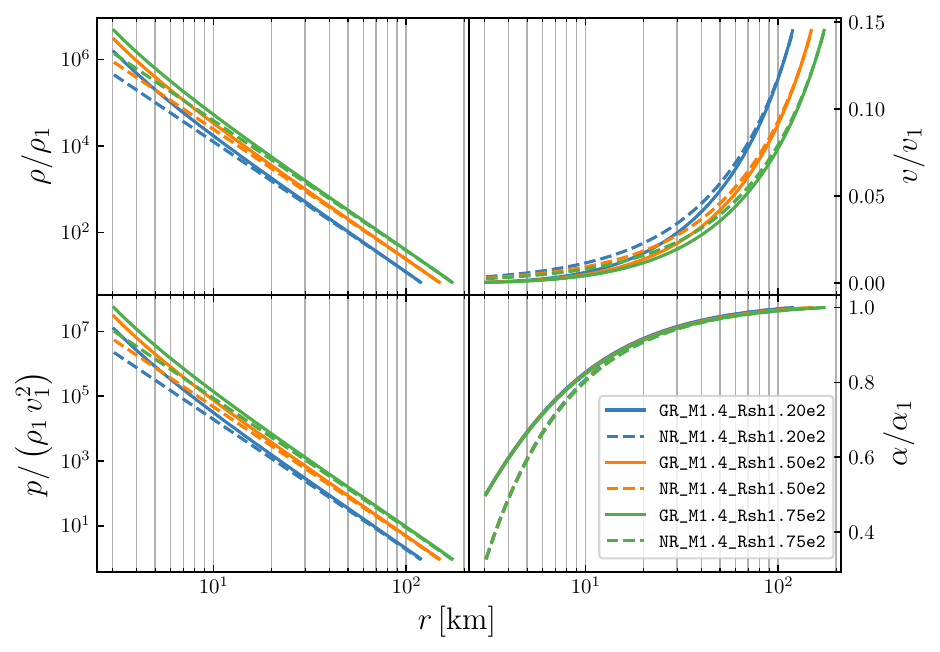}
  \caption{
Post-shock, steady-state solutions for GR (solid) and NR (dashed)
equations as functions of coordinate distance $r\in\left[3\,\km,\rsh\right]$
for three models having the same
accretion rate of 0.3$\,\msun/\s$,
the same PNS mass of 1.4$\,\msun$, and shock radii
120 km (blue), 150 km (orange), and 175 km (green).
For this PNS mass, the Schwarzchild radius is $\sim$1 km.
Quantities defined just ahead of the shock are denoted with a subscript ``1".
The top-left panel shows the comoving baryon mass density
normalized to its value just ahead of the shock,
the top-right panel shows the radial component of the fluid three-velocity
normalized to its value just ahead of the shock,
the bottom-left panel shows the comoving pressure
normalized to the Newtonian ram pressure just ahead of the shock,
$\rho_{1}\,v_{1}^{2}$,
and the bottom right panel shows the lapse function (solid) and
the Newtonian approximation to the lapse function (dashed),
$1+\Phi$, with $\Phi$ the Newtonian gravitational potential, normalized
to their values at the shock.}
  \label{fig.compSS_loXi}
\end{figure*}

For the two lower-compactness cases (see \myeqref{eq.compactness})
\citep[e.g., see][]{bmh2013,mjm2015,brv2020},
\figref{fig.compSS_loXi} shows steady-state accretion shock solutions
as functions of coordinate distance $r$.
(Note that in \figref{fig.compSS_loXi} and \figref{fig.compSS_hiXi} below,
for additional information, the steady-state solutions are plotted from the shock all
the way down to $r=3$~km.
For our numerical simulations, the inner boundary is placed at
$\rpns\in\{20,40\}$~km, which then defines the compactness of our models.)

In general, the magnitudes of the density, velocity,
and pressure just below the shock decrease as the shock radius increases
\citep[e.g., see Eqs. (1-3) in][]{bmd2003}.
From the top-right panel of \figref{fig.compSS_loXi},
it can be seen that the velocities in the GR and NR cases agree
well near the shock and deviate from each other for smaller radii,
with the velocity being smaller in the GR case than in the NR case.
The top-left and bottom-left panels show that the densities and pressures
in the GR case are larger than their NR counterparts at smaller radii.
The slope of the NR density profile
matches expectations of $\rho\left(r\right)\propto r^{-3}$ \citep{bmd2003},
but the GR density profile deviates noticeably
from this as the inner-boundary
is approached.
From the bottom-right panel, it can be seen that the lapse function
and its Newtonian approximation begin to deviate from each other
near $r=40\,\km$,
with the degree of deviation increasing
for smaller radii.

For the two higher-compactness cases
\citep[e.g., see][]{lmt2001,wtj2020},
\figref{fig.compSS_hiXi} shows steady-state accretion shock solutions
as functions of coordinate distance $r$.
The profiles show the same trends as those in \figref{fig.compSS_loXi},
although in this case the trends are more pronounced.
One notable difference is that the location of the largest
deviation in the velocity between the NR and GR cases occurs further in,
near $r=12~\km$.
Another notable feature of both Figures \ref{fig.compSS_loXi} and
\ref{fig.compSS_hiXi} is that the fluid velocity in the GR case is consistently
slower than that in the NR case, in agreement with \citet{kc2022}.
\begin{figure*}[htb!]
  \centering
  \includegraphics[width=0.8\textwidth]%
  {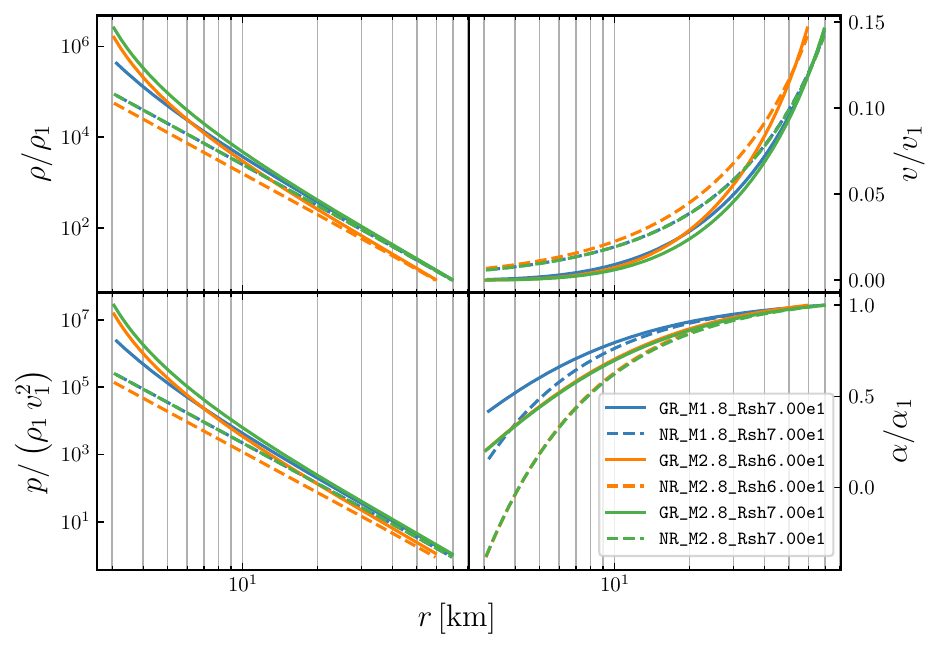}
  \caption{
Post-shock, steady-state solutions for GR (solid) and NR (dashed)
equations as functions of coordinate distance $r\in\left[3\,\km,\rsh\right]$
for three models having the same
accretion rate of 0.3$\,\msun/\s$,
one having a PNS mass of $1.8\,\msun$ and a shock radius of 70 km (blue),
and the other two having PNS mass $2.8\,\msun$
and shock radii 60 km (orange) and 70 km (green).
For the $2.8\,\msun$ PNS models,
the Schwarzchild radius is $\sim$2.1 km.
Quantities defined just ahead of the shock are denoted with a subscript ``1".
The top-left panel shows the comoving baryon mass density
normalized to its value just ahead of the shock,
the top-right panel shows the radial component of the fluid three-velocity
normalized to its value just ahead of the shock,
the bottom-left panel shows the comoving pressure
normalized to the Newtonian ram pressure just ahead of the shock,
$\rho_{1}\,v_{1}^{2}$,
and the bottom right panel shows the lapse function (solid) and
the Newtonian approximation to the lapse function (dashed),
$1+\Phi$, with $\Phi$ the Newtonian gravitational potential, normalized
to their values at the shock.}
  \label{fig.compSS_hiXi}
\end{figure*}

We compare our numerical results with an estimate provided by
\citet{m2020}, which
provides an analytic estimate of the oscillation period
of the SASI, $T_{\mathrm{aa}}$,
based on the assumption that the SASI is an advective-acoustic cycle,
in which a fluid parcel advects from the shock to the PNS surface in time
$\tauad$, which generates acoustic waves that propagate from the
PNS surface to the shock in time $\tauac$.
We modify that formula to include the effects of GR
by including the metric factor, which involves the conformal factor
and which converts the radial coordinate increment to the proper radial
distance increment,
and by replacing the nonrelativistic signal speeds with their relativistic
counterparts,
\begin{equation}
  T_{\mathrm{aa}}\approx\tauad+\tauac
  =\int_{\rsh}^{\rpns}\frac{\sqrt{\gamma_{11}}\,dr}{\lambda^{r}_{0}}
  +\int_{\rpns}^{\rsh}\frac{\sqrt{\gamma_{11}}\,dr}{\lambda^{r}_{+}},
  \label{eq.MullerEst}
\end{equation}
where $\lambda^{r}_{0}$ and $\lambda^{r}_{+}$ are the radial signal speeds of
matter and acoustic waves, respectively.
Using our metric, the signal speeds are \citep{rz2013}
\begin{equation}
  \lambda^{r}_{0}=\alpha\,v^{r}\stackrel{\nr}{=}v^{r},\label{eq.lambda0}
\end{equation}
\begin{widetext}
\begin{equation}
  \lambda^{r}_{+}=\frac{\alpha}{1-v^{i}v_{i}\,c_{s}^{2}}
  \left\{v^{r}\left(1-c_{s}^{2}\right)
  +c_{s}\sqrt{\left(1-v^{i}v_{i}\right)
  \left[\gamma^{11}\left(1-v^{i}v_{i}\,c_{s}^{2}\right)
  -v^{r}\,v^{r}\left(1-c_{s}^{2}\right)\right]}\right\}
  \stackrel{\nr}{=}v^{r}+c_{s},\label{eq.lambda+}
\end{equation}
\end{widetext}
where $c_{s}$ is the sound-speed
and where the second equality in each expression is the nonrelativistic limit.
For 1D problems,
this expression depends only on the steady-state values of $c_{s}$ and $v^{r}$.
We also compare our results with an estimate based on the assumption that
the SASI is a purely acoustic phenomenon.
We define a time, $\tac$, as the time taken by an
acoustic perturbation
to circumnavigate the post-shock cavity
at a characteristic radius $\rac$,
assuming $v^{\theta}\ll c_{s}\,\sqrt{\gamma^{22}}$,
\begin{equation}
  \tac
  :=2\pi\,\rac/\left(h_{\theta}\,\lambda_{+}^{\theta}\right),
  \label{eq.Tac}
\end{equation}
where
\begin{widetext}
\begin{equation}
  \lambda^{\theta}_{+}=\frac{\alpha}{1-v^{i}v_{i}\,c_{s}^{2}}
  \left\{v^{\theta}\left(1-c_{s}^{2}\right)
  +c_{s}\sqrt{\left(1-v^{i}v_{i}\right)
  \left[\gamma^{22}\left(1-v^{i}v_{i}\,c_{s}^{2}\right)
  -v^{\theta}\,v^{\theta}\left(1-c_{s}^{2}\right)\right]}\right\}
  \stackrel{\nr}{=}v^{\theta}+c_{s}/r
\end{equation}
\end{widetext}
is the acoustic wave-speed in the $\theta$-dimension \citep{rz2013} and
$h_{\theta}$ is the scale factor in the $\theta$-dimension.
The expression for $\tac$ is inspired by the Lamb frequency,
which relates the frequency of an acoustic wave to the radius at which it turns
around \citep{hkt2004}.
We delay the specification of $\rac$ until \secref{sec.results}.

\begin{figure*}[htb!]
  \centering
  \includegraphics[width=0.8\textwidth]%
  {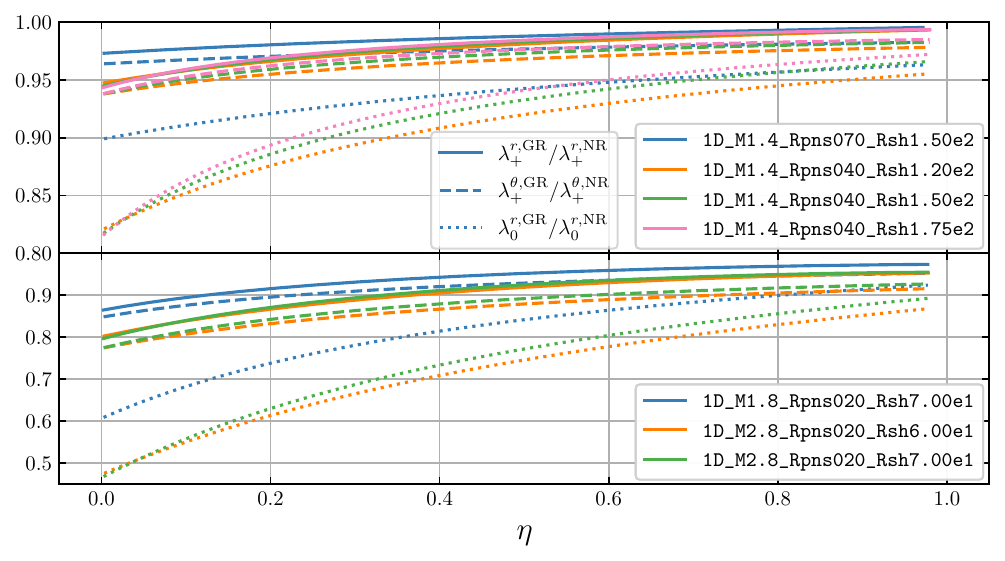}
  \caption{
Ratio of radial GR acoustic signal speed to radial NR acoustic signal speed
(solid),
ratio of angular GR acoustic signal speed to angular NR acoustic signal speed
(dashed), and
ratio of radial GR advective signal speed to radial NR advective signal speed
(dotted),
plotted as functions of $\eta$ for all models.
Lower-compactness models are shown in the top panel
and higher-compactness models
are shown in the bottom panel.
All signal speeds have been multiplied with the appropriate scale factors.}
  \label{fig.wavespeeds}
\end{figure*}

In \figref{fig.wavespeeds}, we plot
$\lambda^{r,     \gr}_{+}/\lambda^{r,     \nr}_{+}$,
$\lambda^{\theta,\gr}_{+}/\lambda^{\theta,\nr}_{+}$, and
$\lambda^{r,     \gr}_{0}/\lambda^{r,     \nr}_{0}$
as functions of $\eta$, defined as
\begin{equation}
  \eta\left(r\right):=\left(r-\rpns\right)/\left(\rsh-\rpns\right),
  \label{eq.eta}
\end{equation}
for all of our models.
In all cases, the signal speeds are slower in the GR case.
This difference is accentuated in the
higher-compactness models and for smaller radii.

The growth rate of the SASI depends on the
steady-state conditions below the shock; however, obtaining an analytic estimate
for this is difficult,
and although there have been efforts to explain the physics governing the
growth rate assuming nonrelativistic models
\citep[e.g.,][]{bm2006,fgs2007,l2007,l2008,f2009,gf2012},
an analytic estimate remains an open question
and no estimate exists for a GR model.
Here, we aim to compare the NR and GR growth rates with numerical simulations.

We emphasize that our goal is to characterize the SASI in terms of its period
and growth rate, and to compare their GR and NR values.
Determining its physical origin, whether it be advective-acoustic or purely
acoustic, is beyond the scope of this study.
The rough estimates provided by \myeqref{eq.MullerEst} and \myeqref{eq.Tac} are
merely intended as points of reference for our numerically determined values.

\section{Simulation Code and Setup}

We perform our simulations with \thornado{}%
\footnote{\url{https://github.com/endeve/thornado}},
an open-source code under development,
aiming to simulate CCSNe.
\thornado{} uses high-order discontinuous Galerkin (DG) methods
to discretize space and strong-stability-preserving Runge--Kutta
(SSP-RK) methods to evolve in time.
For details on the implementation in the nonrelativistic case,
see \citet{ebd2019} and \citet{pbe2021};
for the relativistic case, see \citet{dem2020} and
Dunham et al. (2024) (in prep.).
All of our simulations use the HLL Riemann solver \citep{hll1983},
a quadratic polynomial representation (per dimension) of the
solution in each element, and
third-order SSP-RK methods for time integration \citep{gst2001}
with a timestep
$\Delta t
:=C_{\textsc{cfl}}
\frac{1}{d\left(2\,k+1\right)}
\min\limits_{i\in\left\{1,\ldots,d\right\}}
\Delta x^{i}/\left|\lambda^{i}\right|$,
where $C_{\textsc{cfl}}=0.5$ is the CFL number,
$k$ the degree of the approximating polynomial (in our case, $k=2$),
$\Delta x^{i}$ the mesh width in the $i$\th{} dimension,
$\lambda^{i}$ the fastest signal speed in the $i$\th{} dimension,
and $d$ the number of spatial dimensions \citep{cs2001}.

Two important aspects of successful implementations of
the DG method are mitigating spurious
oscillations and enforcing physical realizability
of the polynomial approximation of the solution.
To mitigate oscillations, \thornado{} uses the \texttt{minmod}
limiter and applies it to the characteristic fields
\citep[see, e.g.,][]{s1987,pbe2021}.
For the interested reader, we set the $\beta_{\textsc{tvd}}$ parameter
for the limiter, defined in \citet{pbe2021}, to $1.75$ for all runs.
\thornado{} also uses the troubled-cell indicator described
in \cite{fs2017} to decide on which
elements to apply the \texttt{minmod} limiter;
for the threshold of that indicator
we use the value $5\times10^{-3}$.
To enforce physical realizability of the solution in the NR case,
\thornado{} uses the positivity limiter described in \citet{zs2010},
and in the GR case, uses the limiter described in \citet{qsy2016};
for the thresholds of both limiters we use the value $10^{-13}$.

The hydrodynamics in \thornado{} has recently been coupled to
\amrex{}\footnote{\url{https://github.com/AMReX-codes/amrex}},
an open-source software library for block-structured adaptive
mesh refinement and parallel computation with MPI \citep{z2019};
however, our simulations are all performed on a uni-level mesh.

Our computational domain, $\mc{D}$, is defined for all models as
$\mc{D}:=\left[\rpns,1.5\,\rsh\left(t=0\right)\right]\times\left[0,\pi\right]$.
The radial extent allows us to determine whether or not the SASI
has become nonlinear, which we define to be when any radial coordinate
of the shock exceeds $10\%$ of the initial shock radius.
All simulations are evolved sufficiently long to achieve ten
full cycles of the SASI.
In some cases, the shock exceeds our threshold of nonlinearity before
completing ten full cycles;
in those cases, we only use data from the linear regime.

The PNS is treated as a fixed, spherically symmetric mass
in order to maintain a steady-state,
and we ignore the self-gravity of the fluid.
Because the largest accretion rate we consider
is $0.5\,\msun\,\s^{-1}$ and because this
lasts for a maximum of $550\,\ms$ in our 2D models,
the most mass that would accrete onto the PNS is $0.275\,\msun$
and therefore would provide
a sub-dominant contribution to $\mpns$ in our simulations.

We consider models with three free parameters:
the mass of the PNS, $\mpns$,
the radius of the PNS, $\rpns$,
and the initial radius of the shock, $\rsh$.
We also varied the mass accretion rate, $\mdot$, but found the
oscillation periods and growth rates to be insensitive to this
parameter, and we do not discuss these models further;
all following discussion
is for models with an accretion rate $\mdot=0.3\,\msun\,\s^{-1}$.
Our choice of parameters is motivated by the physical scale of CCSNe;
the ranges of our parameter space are informed
by models from \citet{lmt2001,bmh2013,mjm2015,brv2020,wtj2020},
and can be found in \tabref{tab.modelParameters}.
We also classify our simulations by their compactness, $\xi$, which we
define as \citep{oo2011},
\begin{equation}
  \xi:=\frac{M/\msun}{\rpns/\left(20\,\km\right)}.
  \label{eq.compactness}
\end{equation}

\begin{deluxetable*}{ccccc}[htb!]
  \tablecaption{Model Parameters}
  \tablehead{ %
  \colhead{Model} &
  \colhead{$\mpns\left[\msun\right]$} &
  \colhead{$\rpns\left[\km\right]$} &
  \colhead{$\rsh\left[\km\right]$} &
  \colhead{$\xi$} }
  \startdata
  \texttt{M1.4\_Rpns070\_Rsh1.50e2} & 1.4 & 70 & 150 & 0.4 \\
  \texttt{M1.4\_Rpns040\_Rsh1.20e2} & 1.4 & 40 & 120 & 0.7 \\
  \texttt{M1.4\_Rpns040\_Rsh1.50e2} & 1.4 & 40 & 150 & 0.7 \\
  \texttt{M1.4\_Rpns040\_Rsh1.75e2} & 1.4 & 40 & 175 & 0.7 \\
  \texttt{M1.8\_Rpns020\_Rsh7.00e1} & 1.8 & 20 &  70 & 1.8 \\
  \texttt{M2.8\_Rpns020\_Rsh6.00e1} & 2.8 & 20 &  60 & 2.8 \\
  \texttt{M2.8\_Rpns020\_Rsh7.00e1} & 2.8 & 20 &  70 & 2.8 \\
  \enddata
  \label{tab.modelParameters}
  \tablecomments{Model parameters chosen for the \nModels{} models.
All models were run with both GR and NR.}
\end{deluxetable*}
In our model naming convention,
we first list whether the model used GR or NR along with the dimensionality
(1D or 2D),
followed by the mass of the PNS in Solar masses, the radius of the PNS
in kilometers,
and lastly the shock radius in kilometers; e.g.,
the 2D GR model with $\mpns=1.4\,\msun$, $\rpns=40\,\km$,
and $\rsh=150\,\km$ is named \texttt{GR2D\_M1.4\_Rpns040\_Rsh1.50e2}.
If we compare an NR model with a GR model having the same parameters
we may drop that specification from the model name.
If no confusion will arise, we may also drop the dimensionality.

The inner radial boundary corresponds to the surface of the PNS.
To determine appropriate inner-boundary conditions in the GR case,
we assume $D$ and $\tau$ follow
power laws in radius; from the initial conditions,
we extrapolate $D$ and $\tau$ in radius
using a least-squares method with data from the innermost five
elements on the grid to determine the appropriate exponents.
The radial momentum density interior to $\rpns$ is
kept fixed to its initial value.
We leave the outer radial boundary values
fixed to their initial values for all fields.
In the polar direction,
we use reflecting boundary conditions at both poles.
For the inner-boundary conditions in the nonrelativistic case, we
assume $\rho\left(r\right)\propto r^{-3}$ and
$E\left(r\right)\propto r^{-4}$,
the latter of which follows from our assumption of $\rho\propto r^{-3}$,
\myeqref{eq.idealEOS}, \myeqref{eq.polytrope}, the assumption of $\Gamma=4/3$,
and the assumption of a small velocity at the PNS surface.

\begin{figure*}[htb!]
  \centering
  \includegraphics[width=0.8\textwidth]%
  {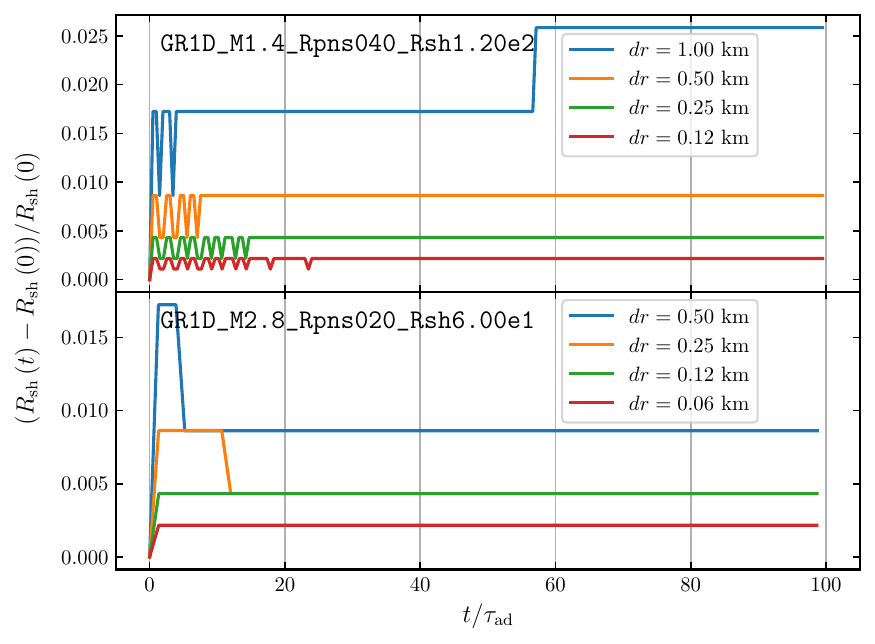}
  \caption{%
Relative deviation of the shock radius from its initial location versus
$t/\tauad$ for different radial resolutions $dr$
for model \texttt{GR1D\_M1.4\_Rpns040\_Rsh1.20e2} (top panel)
and model \texttt{GR1D\_M2.8\_Rpns020\_Rsh6.00e1} (bottom panel).}
  \label{fig.resReq}
\end{figure*}

For the
$\xi\in\left\{0.4,0.7\right\}$ models
we enforce a radial resolution of
0.5 km per element for all runs, which we found necessary for
the shock in an unperturbed model to not deviate by more than 1\%
over 100 advection times.
This is shown in the top panel of \figref{fig.resReq},
which plots the relative deviation of the shock radius from its initial
position as a function of $t/\tauad$
for runs with different radial resolutions
for model \texttt{GR1D\_M1.4\_Rpns040\_Rsh1.20e2}.
These results suggest that the
steady-state is not maintained if the radial resolution is too coarse;
e.g., greater than about 1 km.

For the
$\xi\in\left\{1.8,2.8\right\}$ models
we enforce a radial resolution of
0.25 km per element for all runs
in order to maintain the same radial resolution of the pressure scale height,
$p\left|dp/dr\right|^{-1}$,
as in the $\xi\in\left\{0.4,0.7\right\}$ models,
while also ensuring that the shock does not deviate from its initial
location by more than 1\%.
This can be seen in the bottom panel of \figref{fig.resReq},
which plots the same quantity as the top panel,
but for model \texttt{GR1D\_M2.8\_Rpns020\_Rsh6.00e1}.

To verify that our chosen angular resolution of 64 elements
($\sim$2.8$^{\circ}$) is sufficient to resolve the angular variations
of the fluid, we run two
additional simulations of model \texttt{NR2D\_M2.8\_Rpns040\_Rsh1.20e2},
one with 128 angular elements and one with 256 angular elements.
From those runs, we extract the best-fit values for the growth rates and
oscillation periods (see \secref{sec.results})
and find them to not significantly differ from those of the
64-angular-element run.

\begin{figure*}[htb!]
  \centering
  \includegraphics[width=0.8\textwidth]%
  {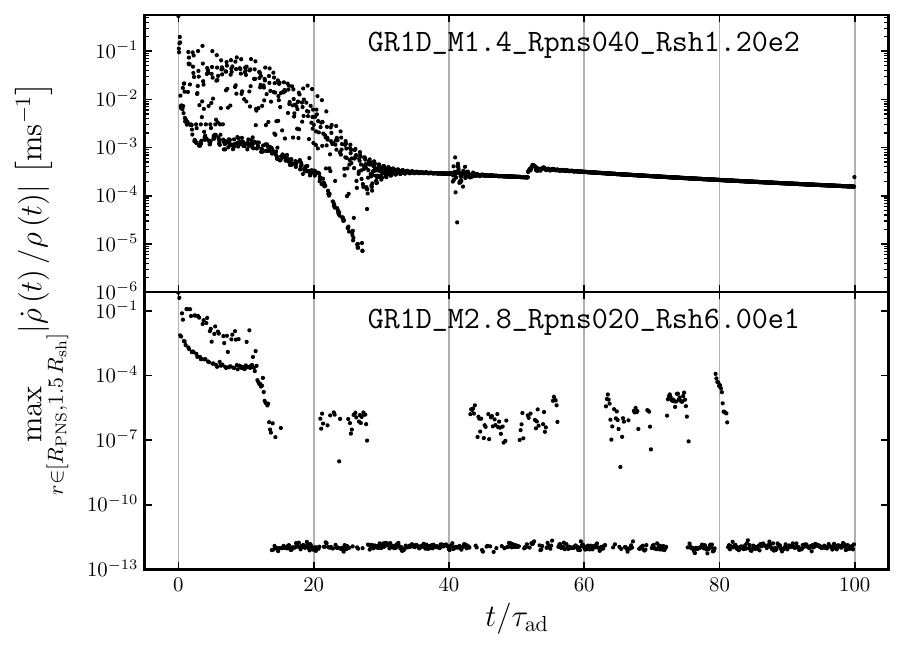}
  \caption{Maximum, at each snapshot, of the absolute value of
the normalized time derivative
of the mass density versus $t/\tauad$ for model
\texttt{GR1D\_M1.4\_Rpns040\_Rsh1.20e2} (top panel)
and model \texttt{GR1D\_M2.8\_Rpns020\_Rsh6.00e1} (bottom panel).}
  \label{fig.relax}
\end{figure*}

Our simulations
are initialized with the steady-state solutions discussed in \secref{ss.ssas},
and we take extra care to minimize initial transients.
The initial conditions we obtain
come from solving Eqs. (\ref{eq.polytrope}), (\ref{IC.eq1}), and
(\ref{IC.eq2}) in the GR case,
and Eqs. (\ref{eq.polytrope}), (\ref{eq.MassConservationNR}), and
(\ref{eq.BernoulliNR}) in the NR case, which are
not exact solutions of our discretized equations,
so transients will be present when simulations are initialized with these
solutions.
To mitigate the effects of the transients,
the fields are set up in 1D with the method described above and then
evolved for 100 advection times, which was experimentally determined to be of
sufficient duration to quell any transients.
We verify that the system has achieved a steady-state
by plotting, in \figref{fig.relax}, the maximum,
at each snapshot, of the absolute value of the normalized
time derivative of the mass density
versus $t/\tauad$ for model \texttt{GR1D\_M1.4\_Rpns040\_Rsh1.20e2}
(top panel)
and model \texttt{GR1D\_M2.8\_Rpns020\_Rsh6.00e1}
(bottom panel); other models exhibit similar behavior.
For example, in the top panel,
it can be seen that the model settles down after approximately
35 advection times,
followed by two slight increases near 45 and 50 advection times,
then settles down until we end the simulation after 100 advection times.
We attribute the two slight increases to limiters activating
when the shock crosses an element boundary.
The relaxed 1D data is mapped to 2D, a
perturbation to the pressure is applied (see below), and the system is
evolved.
The initial relaxation in 1D
removes numerical noise and allows for a smaller perturbation
amplitude, leading to a longer-lasting linear regime, which makes for a
cleaner signal.

We seed the instability by imposing a pressure perturbation onto the
steady-state flow below the shock,
\begin{equation}
  p\left(r,\theta\right)
  =p\left(r\right)+\delta p\left(r,\theta\right),
\end{equation}
with $p\left(r\right)$ the steady-state pressure at radius $r$,
and where $\delta p\left(r,\theta\right)/p\left(r\right)$ is defined in a
scale-independent manner as
\begin{equation}
  \frac{\delta p\left(r,\theta\right)}{p\left(r\right)}
  :=10^{-6}\,\
  \exp\left(\frac{-\left(\eta-\etac\right)^{2}}{2\,\sigma^{2}}\right)\,
  \cos\theta,
\end{equation}
with $\eta$ defined in \myeqref{eq.eta},
and where $\etac=0.75$ and $\sigma=0.05$.
The perturbation is not allowed to extend into the pre-shock flow.
We opted to perturb the post-shock flow as opposed to
the pre-shock flow
after having tried various pre-shock and post-shock perturbations
and finding that pre-shock perturbations generate noise when
they cross the shock front, thus creating a noisy signal and making
it more difficult to extract quantities of interest.
Similarly, we choose a Gaussian profile
because the smoothness of the profile generates less noise than, e.g.,
a top-hat profile.
The $\cos\theta$ factor is meant to excite an $\ell=1$ Legendre mode
of the SASI.
While this perturbation method does not exactly mimic the hydrodynamics
inside a CCSN, it is sufficient to study the SASI
in the linear regime.

\section{Results and Discussion}
\label{sec.results}

Here we discuss our analysis methods and compare the SASI growth
rates and oscillation periods for our \nModels{} models in GR and NR.

\subsection{Analysis Methods}

To extract SASI growth rates from our simulations,
we follow \citet{bm2006} and expand a quantity
affected by the perturbed flow,
$A\left(t,r,\theta\right)$, in Legendre polynomials,
\begin{equation}
  A\left(t,r,\theta\right)
  =\sum\limits_{\ell=0}^{\infty}
  G_{\ell}\left(t,r\right)\,P_{\ell}\left(\cos\theta\right),
\end{equation}
where we normalize the $P_{\ell}$ such that
\begin{equation}
  \int_{-1}^{1}P_{\ell}\left(x\right)\,P_{\ell'}\left(x\right)\,dx
  =\delta_{\ell\ell'},
\end{equation}
with $\delta_{\ell\ell'}$ the Kronecker delta function.
Then,
\begin{align}
  G_{\ell}\left(t,r\right)&:=
  \int_{-1}^{1}A\left(t,r,\theta\right)\,
  P_{\ell}\left(\cos\theta\right)\,d\left(\cos\theta\right)\notag\\
  &=\int_{0}^{\pi}
  A\left(t,r,\theta\right)\,P_{\ell}\left(\cos\theta\right)\,
  \sin\theta\,d\theta.
\end{align}
After experimenting with several quantities,
we decided to use the quantity proposed by \citet{sjf2008},
\begin{equation}
  A\left(t,r,\theta\right)
  :=\frac{1}{\sin\theta}\,\pd{}{\theta}
  \left(v^{\theta}\left(t,r,\theta\right)\,\sin\theta\right),
\end{equation}
where $v^{\theta}$ is the fluid velocity in the polar direction
as measured by an Eulerian observer,
having units of $\rad\,\s^{-1}$.
With $G_{\ell}$, we compute the power in the $\ell$\th{} Legendre mode,
$H_{\ell}\left(t\right)$, by integrating over a shell below the shock,
bounded from below by $r_{a}=0.8\,\rsh$ and from above
by $r_{b}=0.9\,\rsh$,
\begin{align}
  H_{\ell}\left(t\right)&:=
  \int_{0}^{2\pi}\int_{0}^{\pi}
  \int_{r_{a}}^{r_{b}}
  \left[G_{\ell}\left(t,r\right)\right]^{2}
  \left[\psi\left(r\right)\right]^{6}r^{2}\,\sin\theta\,dr\,d\theta\,d\varphi
  \notag\\
  &=4\pi\int_{r_{a}}^{r_{b}}
  \left[G_{\ell}\left(t,r\right)\right]^{2}
  \left[\psi\left(r\right)\right]^{6}r^{2}\,dr.
\end{align}
For the Newtonian runs, $\psi=1$.
We experimented with different values of $r_{a}$ and $r_{b}$
and found that a thin shell just below the shock gave the cleanest signal.

To extract the SASI growth rate and oscillation period from $H_{1}$,
we begin by fitting the simulation data to the function \citep{bm2006}
\begin{equation}
  F_{1}\left(t\right)
  =F\,e^{2\,\omega\,t}\,
  \sin^{2}\left(\frac{2\pi}{T}\,t+\delta\right),
  \label{eq.H1Fit}
\end{equation}
where $\omega$ is the growth rate of the SASI, $T$ is the
oscillation period of the SASI,
$F$ is an amplitude, and $\delta$ is a phase offset.

We fit the data to the model using the Levenberg--Marquardt nonlinear
least squares method \citep[e.g., see][]{m1978},
provided by \scipy 's \curvefit{} function,
which also provides an estimate on the uncertainty of the fit via the diagonal
entries of the covariance matrix; we use this to define the uncertainty
in the growth rate.
The temporal extent over which we perform the fit is defined to begin
after one SASI oscillation and to end after seven SASI oscillations,
where, for simplicity, we use \myeqref{eq.MullerEst} to define the period of one
SASI oscillation.

The period we report is obtained by performing a Fourier analysis:
We integrate the lateral flux in the radial direction,
\begin{equation}
  F^{r}_{\theta}=\sqrt{\gamma}\,\rho\,v^{r}\,v_{\theta}\times\alpha\,h\,W^{2}
  \stackrel{\nr}{=}\sqrt{\gammabar}\,\rho\,v^{r}\,v_{\theta},
  \label{eq.F12}
\end{equation}
from $r_{a}$ to $r_{b}$ (defined as in the computation of the $H_{\ell}$),
integrate the result over
$\bb{S}^{2}:=\left\{\left(\theta,\varphi\right)\in\bb{R}^{2}
\big|\theta\in\left[0,\pi\right],\varphi\in\left[0,2\pi\right)\right\}$,
and then take the Fourier transform of that result
using the \texttt{fft} tool from \scipy.
From the result, $\wt{F}^{r}_{\theta}(\wt{T})$,
we define the period of the SASI as the value of $\wt{T}$ corresponding to the
peak of the Fourier amplitudes, and
we define the uncertainty in the period as the
full width at half maximum (FWHM) of the Fourier amplitudes.
The FFT is computed over the same time interval as the aforementioned fit.
(We do not use the period returned from \curvefit{} because,
for some models, pollution from
higher-order modes spoils the ability of our fitting function to capture
the $\ell=1$ mode.)

\subsection{Overall Trends}

Here we discuss trends that appear across our models.
We summarize our results in \tabref{tab.results},
which lists the model, the best-fit oscillation period and uncertainty,
the best-fit growth rate and uncertainty,
the product of the best-fit growth rate and best-fit oscillation period,
the period assuming an advective-acoustic mechanism, $\taa$
(estimated using \myeqref{eq.MullerEst}),
and the period assuming a purely acoustic mechanism, $\tac$
(estimated using \myeqref{eq.Tac}).
When using \myeqref{eq.Tac}, we choose the characteristic radius $\rac$
to be $0.85\,\rsh$, which is the midpoint of the shell in which we compute
the power.

\begin{deluxetable*}{cccccc}[ht]
  \tablecaption{Results}
  \tablehead %
  { %
  \colhead{Model}                                          &
  \colhead{$T\pm\Delta T\,\left[\ms\right]$}               &
  \colhead{$\omega\pm\Delta\omega\,\left[\ms^{-1}\right]$} &
  \colhead{$\omega\,T$}                                    &
  \colhead{$\taa\,\left[\ms\right]$}                       &
  \colhead{$\tac\,\left[\ms\right]$}                       %
  }
  \startdata
  \texttt{NR\_M1.4\_Rpns070\_Rs1.50e2} & 25.5566 $\pm$ 21.1850 & 0.0631 $\pm$ 0.0011 & 1.6123 & 20.9559 & 36.6791 \\
  \texttt{GR\_M1.4\_Rpns070\_Rs1.50e2} & 30.3854 $\pm$ 21.5598 & 0.0603 $\pm$ 0.0009 & 1.8313 & 22.6462 & 37.4293 \\
  \texttt{NR\_M1.4\_Rpns040\_Rs1.20e2} & 24.6398 $\pm$ 15.6904 & 0.0733 $\pm$ 0.0011 & 1.8054 & 20.8348 & 26.2646 \\
  \texttt{GR\_M1.4\_Rpns040\_Rs1.20e2} & 27.2758 $\pm$ 16.6796 & 0.0578 $\pm$ 0.0011 & 1.5758 & 23.6236 & 26.9372 \\
  \texttt{NR\_M1.4\_Rpns040\_Rs1.50e2} & 37.3656 $\pm$ 23.4864 & 0.0409 $\pm$ 0.0005 & 1.5291 & 34.3279 & 36.6791 \\
  \texttt{GR\_M1.4\_Rpns040\_Rs1.50e2} & 42.9166 $\pm$ 22.4149 & 0.0360 $\pm$ 0.0004 & 1.5461 & 38.6747 & 37.4293 \\
  \texttt{NR\_M1.4\_Rpns040\_Rs1.75e2} & 51.5463 $\pm$ 29.2280 & 0.0300 $\pm$ 0.0003 & 1.5468 & 47.7212 & 46.0841 \\
  \texttt{GR\_M1.4\_Rpns040\_Rs1.75e2} & 58.6525 $\pm$ 28.1381 & 0.0265 $\pm$ 0.0003 & 1.5540 & 53.5372 & 46.8924 \\
  \texttt{NR\_M1.8\_Rpns020\_Rs7.00e1} & 9.8537 $\pm$ 5.1860 & 0.1524 $\pm$ 0.0039 & 1.5016 & 9.2361 & 10.3138 \\
  \texttt{GR\_M1.8\_Rpns020\_Rs7.00e1} & 12.0565 $\pm$ 4.6944 & 0.1031 $\pm$ 0.0022 & 1.2429 & 12.6159 & 10.9053 \\
  \texttt{NR\_M2.8\_Rpns020\_Rs6.00e1} & 6.1939 $\pm$ 3.7227 & 0.2910 $\pm$ 0.0085 & 1.8027 & 5.2343 & 6.5656 \\
  \texttt{GR\_M2.8\_Rpns020\_Rs6.00e1} & 7.5296 $\pm$ 2.5224 & 0.1365 $\pm$ 0.0042 & 1.0282 & 8.6607 & 7.2635 \\
  \texttt{NR\_M2.8\_Rpns020\_Rs7.00e1} & 7.9137 $\pm$ 4.2711 & 0.1897 $\pm$ 0.0054 & 1.5014 & 7.4177 & 8.2693 \\
  \texttt{GR\_M2.8\_Rpns020\_Rs7.00e1} & 10.2013 $\pm$ 3.0563 & 0.0954 $\pm$ 0.0026 & 0.9735 & 12.1099 & 9.0178 \\
  \enddata
  \label{tab.results}
  \tablecomments{
Oscillation periods, growth rates, and their uncertainties for all
fourteen models having the same accretion rate of $0.3\,\msun\,\s^{-1}$.
The uncertainties for the growth rates are defined as the square roots of the
diagonal entries of the covariance matrix corresponding to the growth rate.
The uncertainies for the oscillation period are defined as the full-width
half-maximum values of the Fourier amplitudes (see \figref{fig.fft}).
The fourth column shows the product of the best-fit growth rate multiplied
by the best-fit oscillation period.
The fifth column shows the estimate for the period assuming an
advective-acoustic origin of the SASI (\myeqref{eq.MullerEst}),
and the sixth column shows the estimate for the period assuming
a purely acoustic origin of the SASI (\myeqref{eq.Tac}), where we use
$\rac=0.85\,\rsh$, the midpoint
of the shell in which we compute the power (see \secref{sec.results}).}
\end{deluxetable*}

As a first general trend, we see that, for a given PNS radius,
the oscillation period increases as the shock radius increases,
as seen clearly in the second column of \tabref{tab.results}.
This is expected, for as the shock radius increases,
the waves supported by the fluid must traverse a larger region,
therefore each cycle will take longer.
As a second general trend, we observe that,
for a given PNS radius, the growth rate decreases as the shock radius
increases.

\subsection{Impact of GR}

Next we discuss the impact of GR on the oscillation period and the growth rate.

\subsubsection{Oscillation Period}

\begin{figure*}[htb!]
  \centering
  \begin{minipage}{\textwidth}
    \begin{minipage}{0.5\textwidth}
      \includegraphics[width=\textwidth]%
      {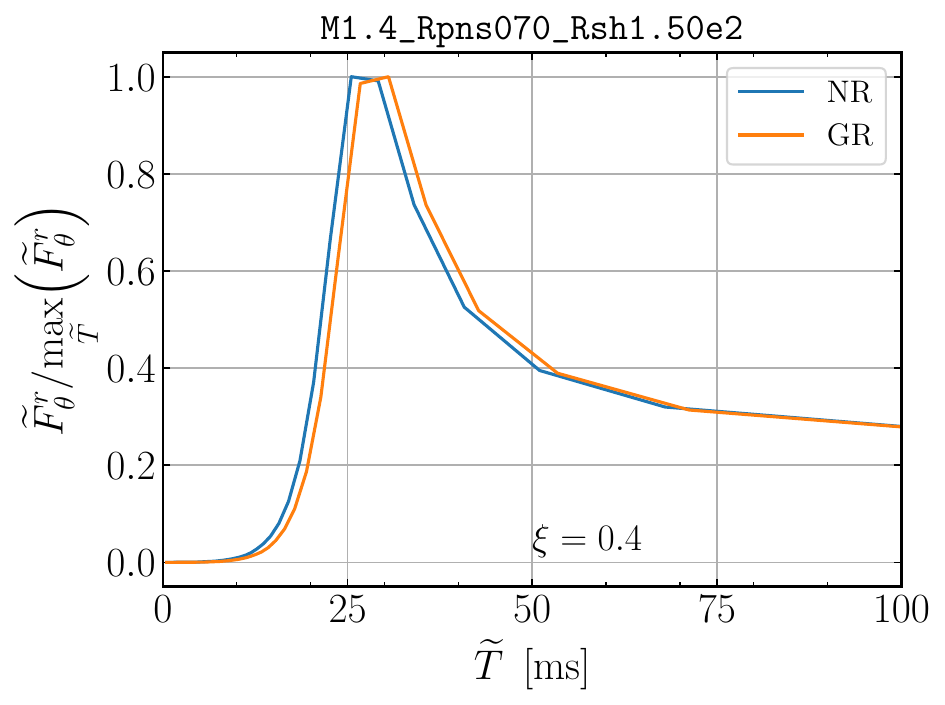}
    \end{minipage}
    \hfill
    \begin{minipage}{0.5\textwidth}
      \includegraphics[width=\textwidth]%
      {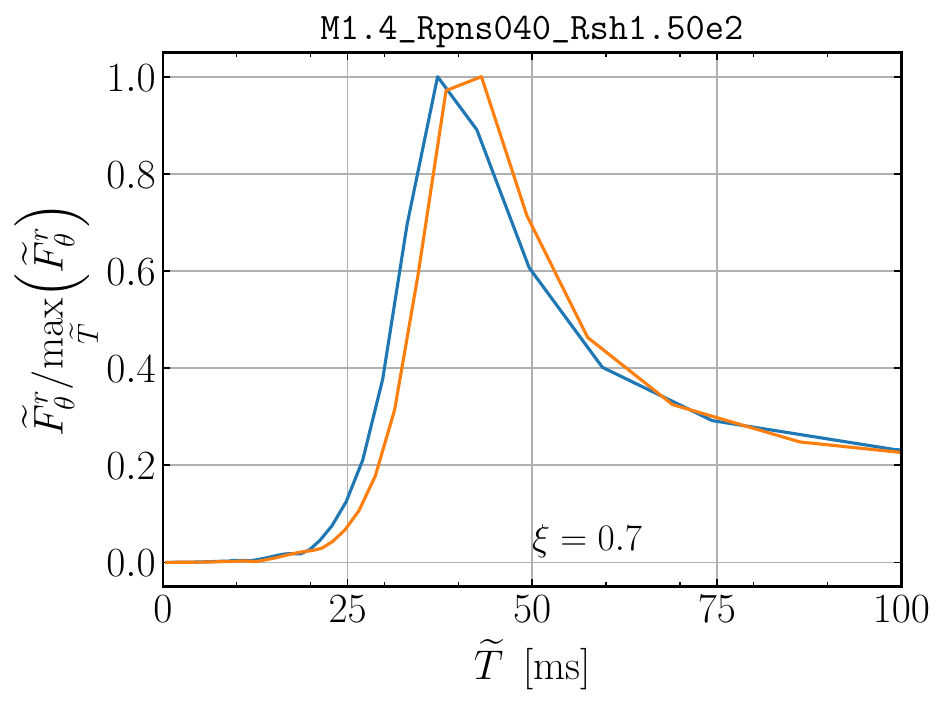}
    \end{minipage}
    \vfill
    \begin{minipage}{0.5\textwidth}
      \includegraphics[width=\textwidth]%
      {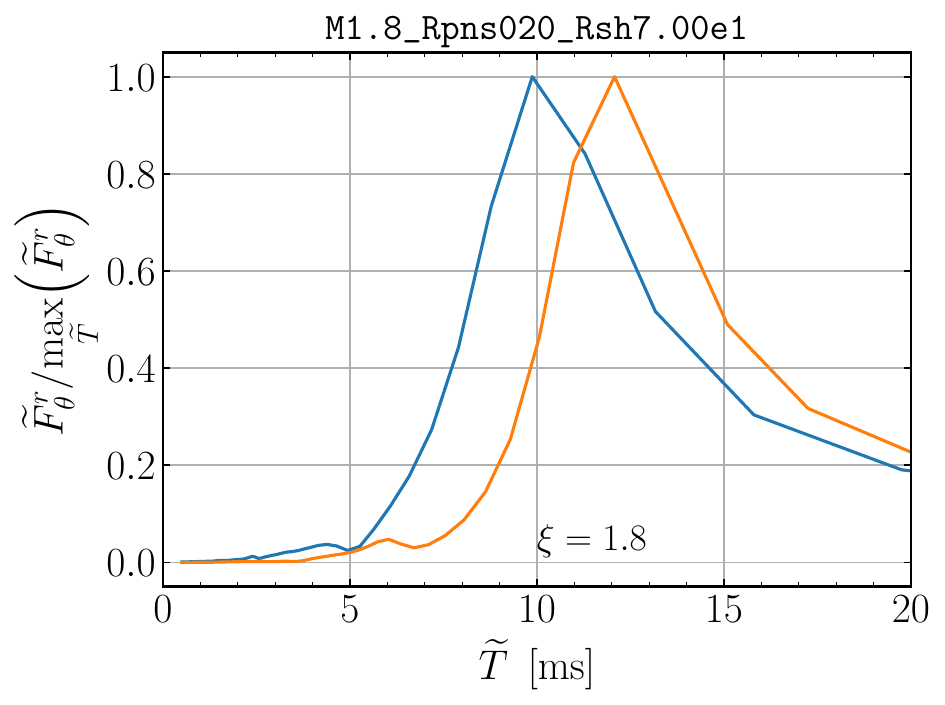}
    \end{minipage}
    \hfill
    \begin{minipage}{0.5\textwidth}
      \includegraphics[width=\textwidth]%
      {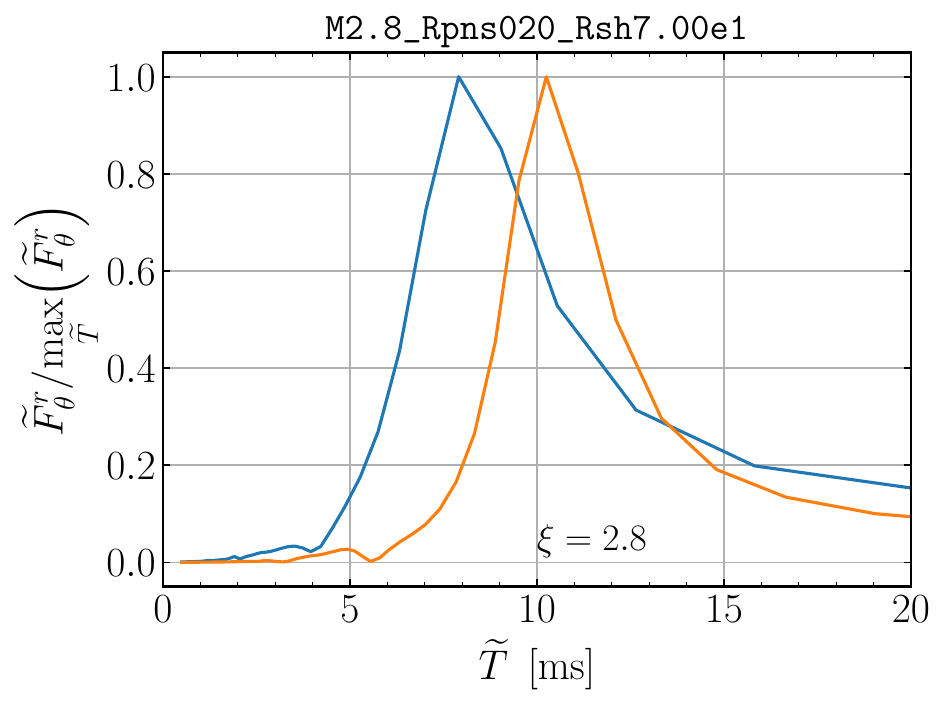}
    \end{minipage}
  \end{minipage}
  \caption{
Amplitudes of Fourier transforms of \myeqref{eq.F12},
normalized by their largest values,
as functions of $\wt{T}$ in $\ms$,
where $\wt{T}$ is the inverse of the frequency returned by the FFT.
The compactness of each model is shown in the bottom of the respective panels.
NR results are shown in blue and GR results are shown in orange.}
  \label{fig.fft}
\end{figure*}

First we discuss how the oscillation period varies with
PNS compactness and shock radius.
In \figref{fig.fft}, we plot the
amplitudes of the Fourier transform of \myeqref{eq.F12}, $\wt{F}^{r}_{\theta}$,
versus $\wt{T}$ for four models,
where $\wt{T}$ is defined as the inverse of the
frequency determined by the FFT.
No windowing was applied when computing the FFT of the signal.
We see that the difference in the optimal period, $T$
(defined as the $\wt{T}$ associated with the largest Fourier amplitude),
between NR and GR increases with increasing compactness,
with the GR period consistently longer than the NR period.
This can be explained by differences in the structure of the post-shock
solutions; in particular, the signal speeds, which are shown in
\figref{fig.wavespeeds}.
Both the radial acoustic and advective signal speeds,
as well as the angular acoustic signal speed,
are consistently smaller in GR.
Because of this, the period is longer for a given model
when GR is used, regardless of whether the SASI is governed by an
advective-acoustic or a purely acoustic cycle.

\begin{figure*}[htb!]
  \centering
  \includegraphics[width=0.8\textwidth]%
  {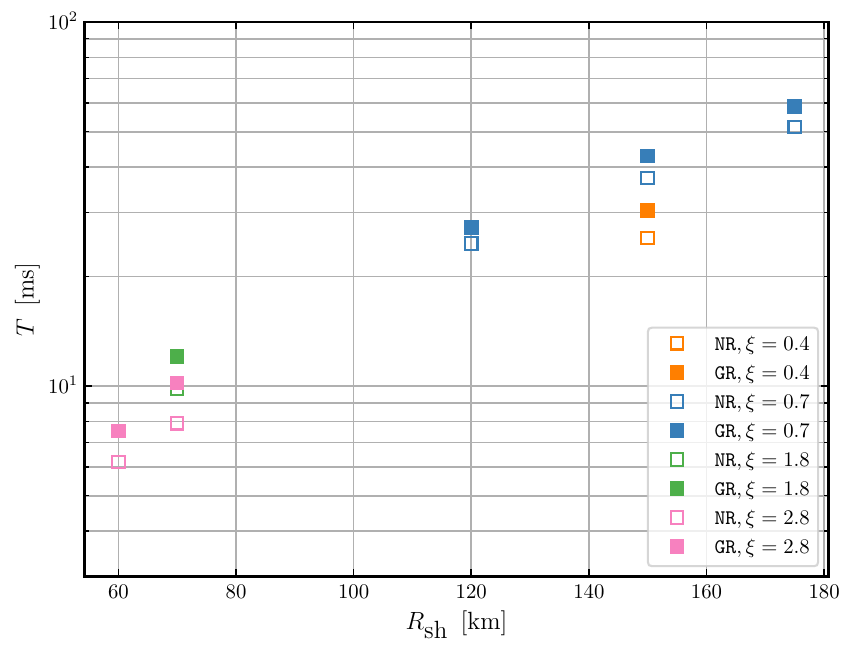}
  \caption{
SASI oscillation period in $\ms$ as a function of
PNS compactness (dimensionless) and
shock radius in $\km$, as determined by the FFT (see text for details).
NR results are shown in open squares
and GR results are shown in solid squares,
with different colors corresponding to different compactnesses.}
  \label{fig.PC}
\end{figure*}

In \figref{fig.PC}, we plot the
SASI period, as provided by the Fourier analysis,
versus initial shock radius for all of our models.
The NR and GR models follow the same general trends.
For $\xi=0.4,0.7,1.8$, and $2.8$,
we find that the ratio $T_{\gr}/T_{\nr}$ is
\PeriodRatioGRoverNRxiA,
$\leq\PeriodRatioGRoverNRxiB$,
\PeriodRatioGRoverNRxiC,
and $\leq\PeriodRatioGRoverNRxiD$, respectively.

For the $\xi=0.4,0.7,1.8$, and $2.8$ models,
the relative differences between $T_{\gr}$ and the advective-acoustic
estimate, \myeqref{eq.MullerEst}, are
\RelDiffACxiA,
$<\RelDiffACxiB$,
\RelDiffACxiC,
and $<\RelDiffACxiD$, respectively.
For those same models,
the relative differences between $T_{\gr}$ and the purely acoustic
estimate, \myeqref{eq.Tac}, are
\RelDiffAAxiA,
$<\RelDiffAAxiB$,
\RelDiffAAxiC,
and $<\RelDiffAAxiD$, respectively.

\subsubsection{Growth Rate}

\begin{figure*}[htb!]
  \centering
  \begin{minipage}{\textwidth}
    \begin{minipage}{0.5\textwidth}
      \includegraphics[width=\textwidth]%
      {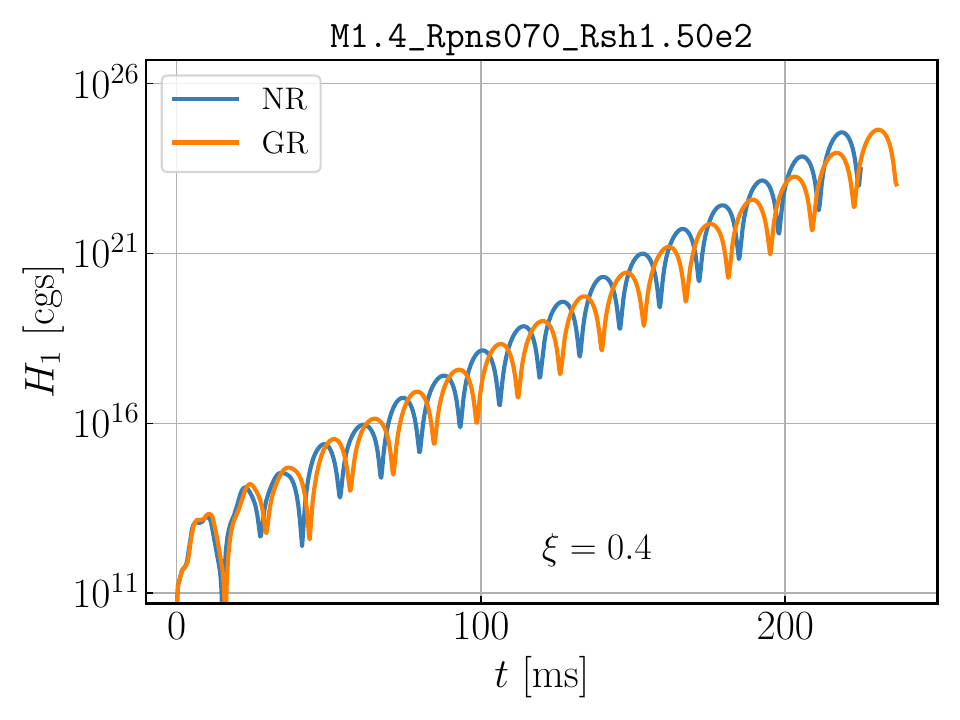}
    \end{minipage}
    \begin{minipage}{0.5\textwidth}
      \includegraphics[width=\textwidth]%
      {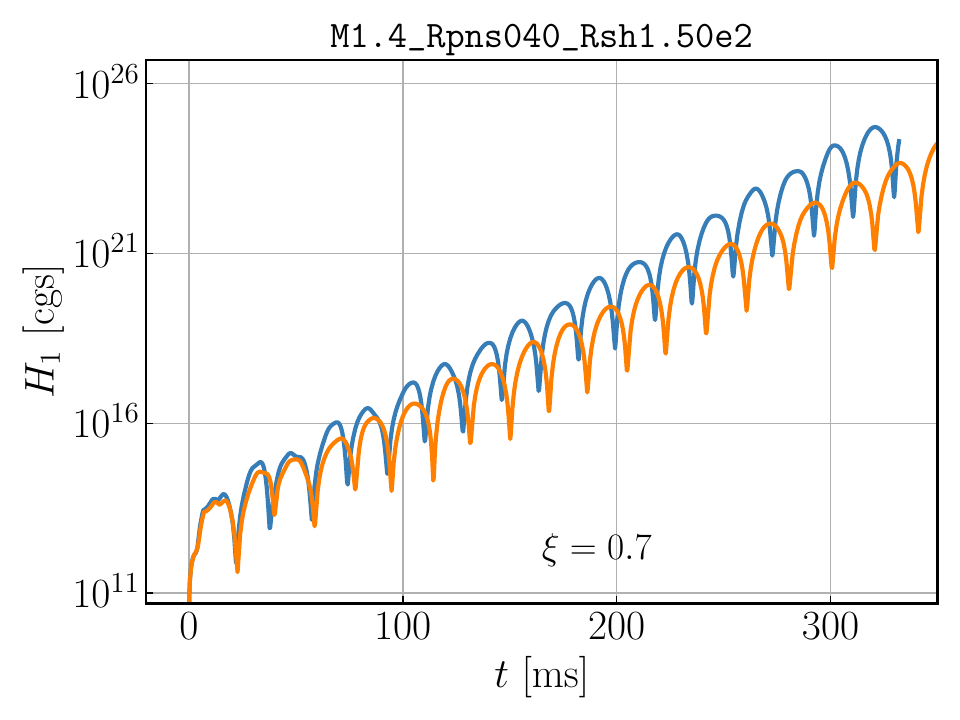}
    \end{minipage}
    \begin{minipage}{0.5\textwidth}
      \includegraphics[width=\textwidth]%
      {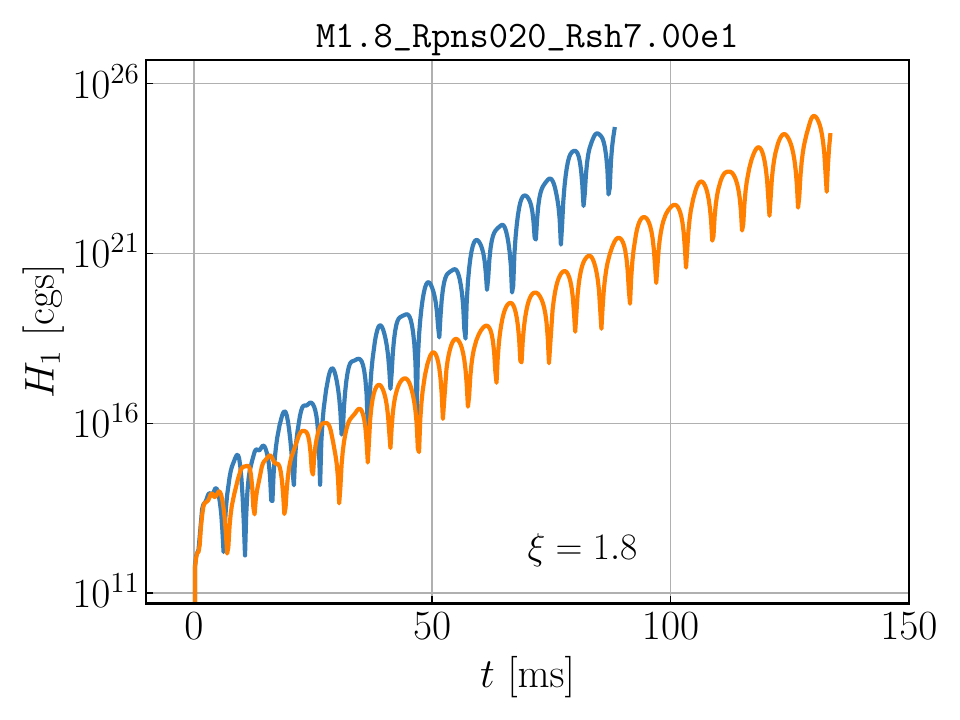}
    \end{minipage}
    \begin{minipage}{0.5\textwidth}
      \includegraphics[width=\textwidth]%
      {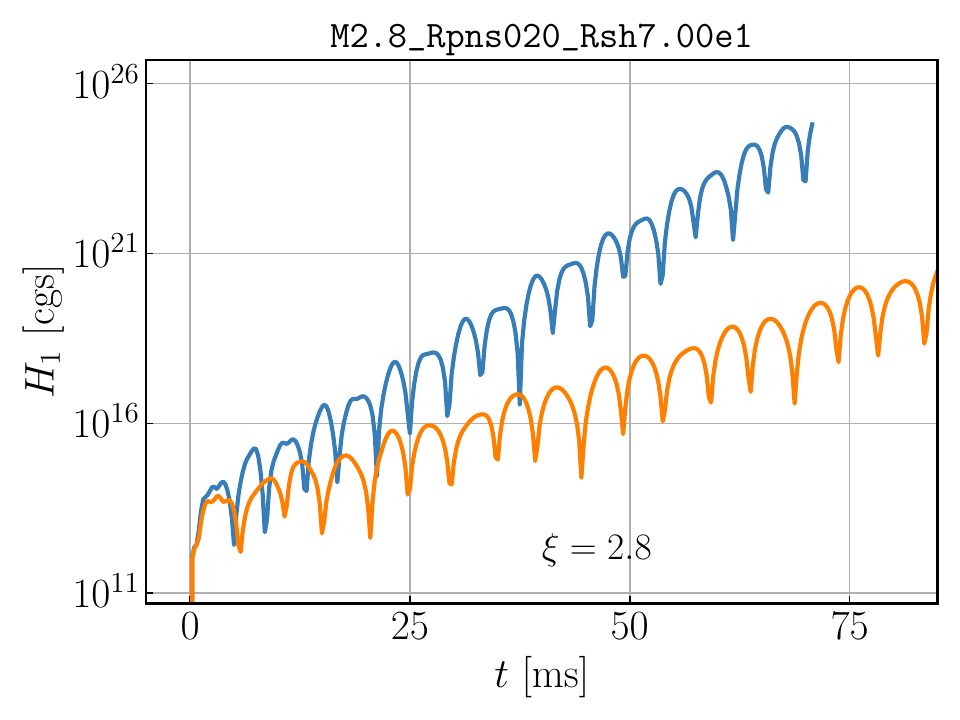}
    \end{minipage}
  \end{minipage}
  \caption{
$H_{1}$ in cgs units versus $t$ in $\ms$.
The layout of the panels is the same as in \figref{fig.fft}.
NR results are shown in blue and GR results are shown in orange.
Note that the horizontal axis limits are different
in each panel.}
  \label{fig.H1}
\end{figure*}

Next we discuss how the growth rate varies with PNS compactness
and shock radius.
In \figref{fig.H1}, we plot the power in the first Legendre mode, $H_{1}$,
during the linear regime,
as a function of time in milliseconds for
the same models as in \figref{fig.fft}.
For all models displayed here, the shock deviates from spherical symmetry by
less than 10\%,
which we consider sufficient for
characterizing the evolution as
being in the linear regime.

\begin{figure*}[htb!]
  \centering
  \includegraphics[width=0.8\textwidth]%
  {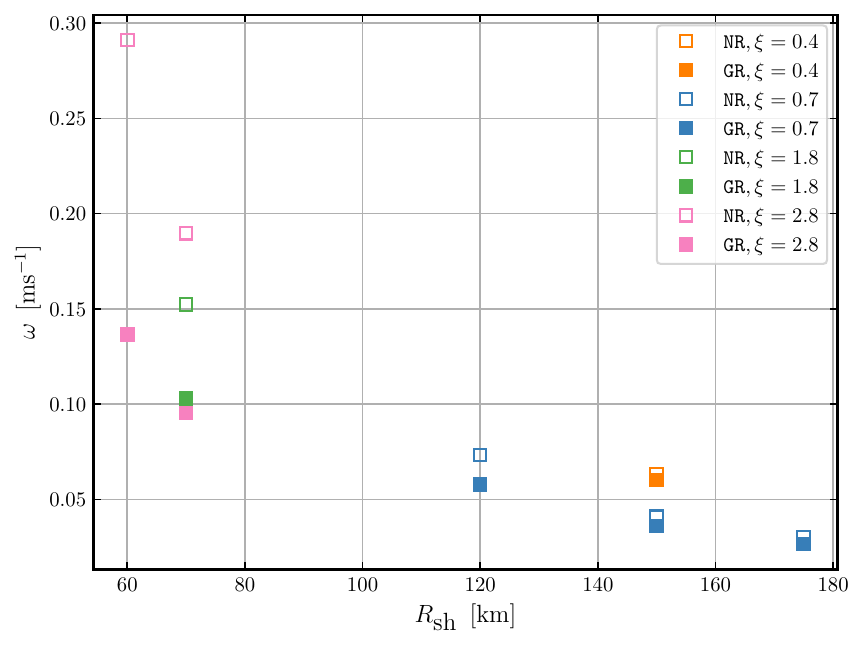}
  \caption{
SASI growth rate in $\ms^{-1}$
as a function of shock radius in $\km$,
as determined by the fit to \myeqref{eq.H1Fit} (see text for details).
NR results are shown in open squares
and GR results are shown in solid squares.}
  \label{fig.GRC}
\end{figure*}

We see a clear trend in the growth rate: the GR models display a slower SASI
growth rate when compared to their NR counterparts.
In \figref{fig.H1}, it can be seen that the growth rate for the $\xi=0.4$
model is practically unaffected by GR,
while all the larger compactness models show non-negligible effects of GR.
This effect is a function of shock radius,
with a smaller shock radius leading to a larger difference
in the NR and GR growth rates.
This effect is drastically enhanced when
looking at the $\xi=2.8$ models.
In these models,
the power in the $\ell=1$ mode is some five
orders of magnitude lower in the GR case
by the end of the simulation (see \figref{fig.H1}),
which is the result of a factor of two difference
in the growth rate along with exponential growth for approximately
ten SASI cycles.
The effects of GR can also be seen in \figref{fig.GRC},
which plots the growth rate for all models as a function of $\rsh$.
Again, in all cases, the growth rate is slower for the GR models and
the difference between GR and NR growth rates increases with
decreasing shock radius and increasing PNS mass -- i.e., under more relativistic
conditions -- the ratio becoming as small as \GrowthRateRatioGRoverNRxiD.

\begin{figure*}[htb!]
  \centering
  \begin{minipage}{\textwidth}
    \begin{minipage}{0.5\textwidth}
      \includegraphics[width=\textwidth]%
      {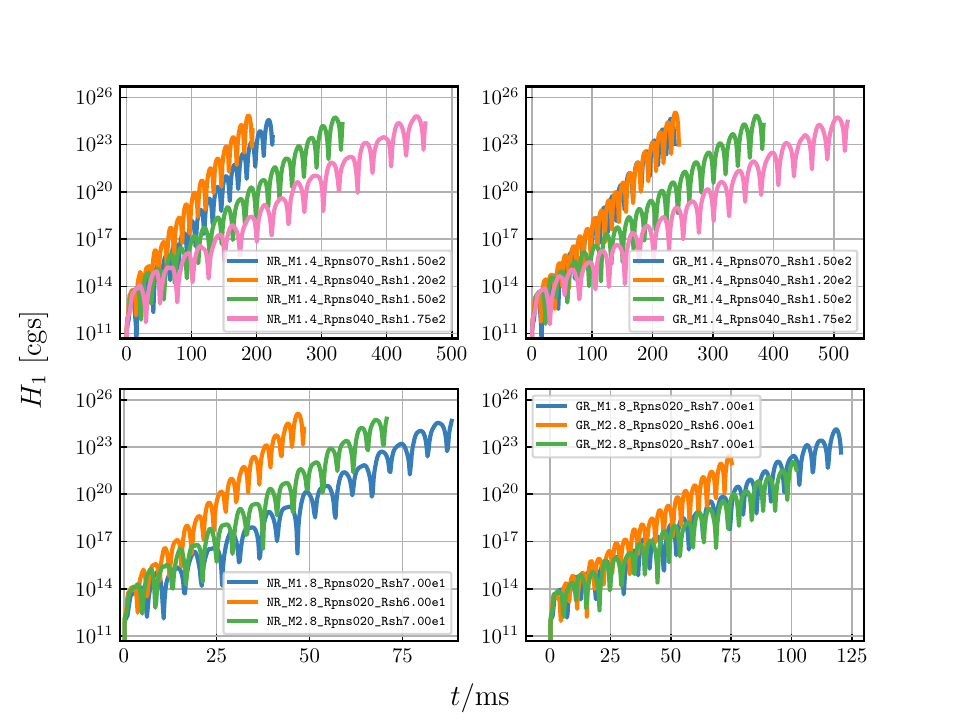}
    \end{minipage}
    \hfill
    \begin{minipage}{0.5\textwidth}
      \includegraphics[width=\textwidth]%
      {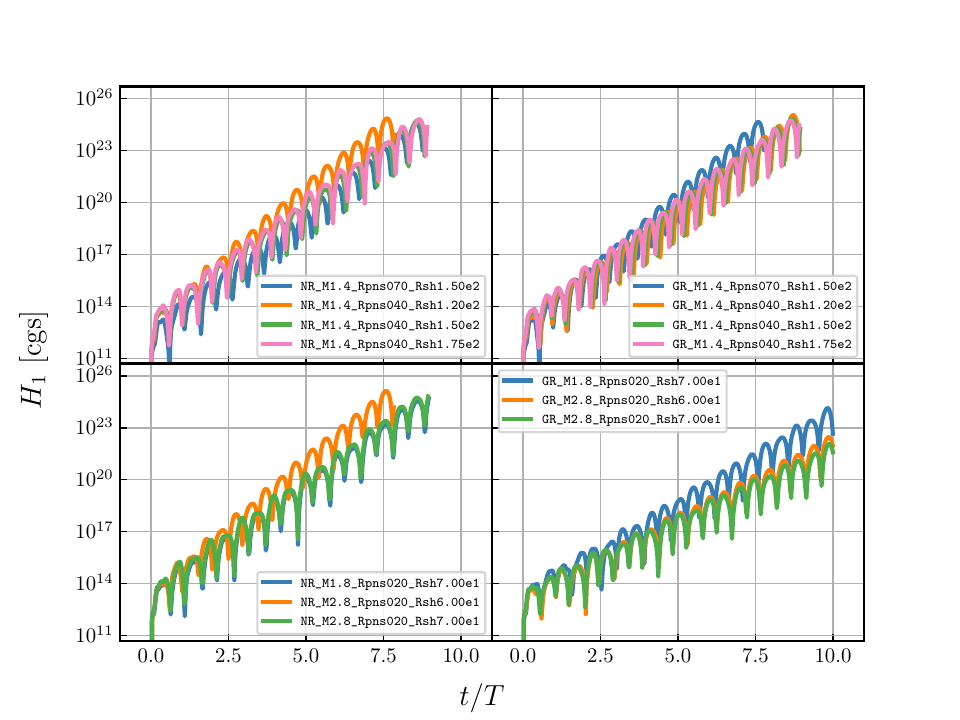}
    \end{minipage}
  \end{minipage}
  \caption{
$H_{1}$ in cgs units versus
$t/\ms$ (left) and versus $t/T$ (right).
In each four-panel figure, the
top-left panel shows the
$\xi\in\left\{0.4,0.7\right\}$, NR models;
the top-right panel shows the
$\xi\in\left\{0.4,0.7\right\}$, GR models;
the bottom-left panel shows the
$\xi\in\left\{1.8,2.8\right\}$, NR models;
and the bottom-right panel shows the
$\xi\in\left\{1.8,2.8\right\}$, GR models.}
  \label{fig.H1t}
\end{figure*}

In \figref{fig.H1t}, we plot $H_{1}$
as a function of $t$ for all models (left panels)
and as a function of $t/T$ for all models (right panels).
Note that for an exponentially growing $\ell=1$ mode,
$H_{1}\propto\exp(\omega t)$;
the gain in one cycle is $|\mathcal{Q}|=\exp(\omega T)$
\citep[e.g., see][]{janka2017}.
Here, we simply refer to $\omega T=\ln|\mathcal{Q}|$ as the SASI efficiency.
(When interpreting the right panels,
note that $T$ is different for each model.)
These figures demonstrate that the SASI cycle efficiency
is approximately constant for a given physical model (NR or GR) and compactness.
This can be seen by, for example, first looking at the top-left panel of the left
figure in \figref{fig.H1t} and noting that the curves all grow
at different rates and then comparing this to the top-left panel of the right
figure, where the curves are brought much closer together when plotted against $t/T$.
Further, all the NR models and the $\xi\in\left\{0.4,0.7\right\}$ GR models
reach about the same total power of $\sim$10$^{25}$-$10^{26}$
after ten oscillations,
while the $\xi\in\left\{1.8,2.8\right\}$
GR models reach a maximum of $\sim$10$^{23}$,
demonstrating that GR modifies both the oscillation period and the growth rate,
but modifies them in such a way as to keep
the SASI efficiency roughly constant for
each of these groupings of models.

\begin{figure*}[htb!]
  \centering
   \includegraphics[width=0.8\textwidth]%
      {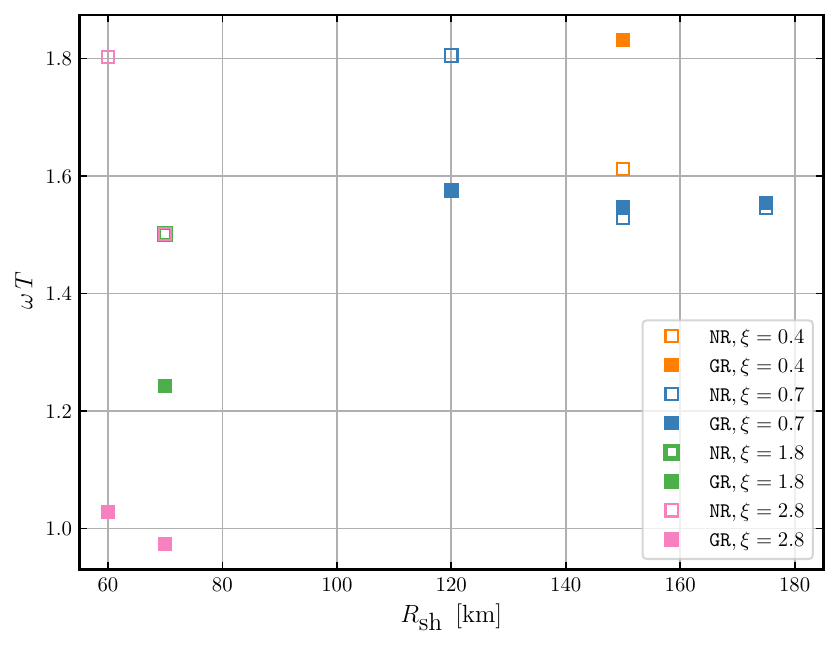}
  \caption{
SASI cycle efficiency
as a function of shock radius in $\km$.
NR results are shown in open squares
and GR results are shown in solid squares.}
  \label{fig.efficiency}
\end{figure*}

The SASI efficiency is further illustrated in
\figref{fig.efficiency}, which plots the efficiency for all of our models.
It can be seen that the SASI efficiency of the NR models takes on values between 1.5 and 1.8,
regardless of compactness.
The GR models follow the same trend for lower compactnesses, but as the
compactness increases the SASI efficiency drops.

\section{Summary, Conclusion, and Future Work}
\label{ss.scfw}

We examined the effect of GR on \nModels{}
idealized, axisymmetric models
of the standing accretion shock instability (SASI) by performing a
parameter study in  which we systematically varied the initial shock radius,
along with the mass and  radius of the proto-neutron star (PNS);
i.e., the compactness of the PNS.
We  compared these runs by measuring the growth rate and oscillation period of
the SASI for all models, which were run once with NRHD and Newtonian gravity
and once  with GRHD and GR gravity.
We set up our simulations to excite a clean $\ell=1$ Legendre mode,
computed the power of the  SASI in that mode within a thin shell just below
the shock, and, following \citet{bm2006}, fit the resulting signal to
\myeqref{eq.H1Fit}.
From the fits, we computed the growth rates and their uncertainties,
and using the FFT,
we computed the oscillation periods and their uncertainties.

For all models, we found that the period of the SASI in the GR case is longer
than that of the SASI in the NR case.
For the $\xi=0.4,0.7,1.8$, and $2.8$ models,
we find ratios, $T_{\gr}/T_{\nr}$, of
\PeriodRatioGRoverNRxiA,
$\leq\PeriodRatioGRoverNRxiB$,
\PeriodRatioGRoverNRxiC,
and $\leq\PeriodRatioGRoverNRxiD$, respectively.
We explained these differences as resulting from differences in the post-shock
flow structure for the GR and NR setups.
For all models, we found that the
growth rate is slower when GR is used,
and significantly slower for the
$\xi=1.8$ and $\xi=2.8$ models,
with the ratio between the GR and NR cases, $\omega_{\gr}/\omega_{\nr}$, as low as
\GrowthRateRatioGRoverNRxiD.
We found that both the growth rate and the oscillation period are practically
independent of the accretion rate for the range of parameter space we
considered.

The connection between our results and the results from realistic
CCSN simulations can be made by considering the trends \emph{across all of the
models considered here}.
First, our results suggest that CCSN simulations based on Newtonian
hydrodynamics and Newtonian gravity may fail to predict correctly the growth
rate of the SASI and its period of oscillation as conditions below the shock
become increasingly relativistic; i.e., increased PNS compactness
and decreased shock radius.
Under such conditions, the growth rate in the Newtonian case may be
overestimated by a factor of two, or more, relative to the GR
case.
Additionally, the period may be underestimated by about 20\%.
Second, as the conditions we considered became increasingly relativistic,
the SASI growth rate continued to increase (\figref{fig.GRC}) and its period continued to decrease (\figref{fig.PC}).
Thus, our studies provide theoretical support for the conclusions reached in
the study of the SASI development in higher-mass progenitors
\citep[see, e.g.,][]{hmw2013,mat2022}, where the time scales for the development
of convection may be long and where the SASI, able to develop on shorter time
scales, is able to provide support to the stalled shock,
potentially sustaining neutrino heating, and, in turn, the development of
neutrino-driven convection, with all of their anticipated benefits for
generating an explosion.

Given the substantial impact of GR we find in our simulations,
we stress the importance of GR treatments in any future studies
aiming to understand the SASI in a CCSN context.

Our results also suggest that CCSN simulations based on Newtonian hydrodynamics
and Newtonian gravity may fail to predict correctly the emission of
gravitational waves by the SASI (specifically, its frequency and amplitude),
a primary target of gravitational wave astronomy given its anticipated existence
in that part of frequency space where current-generation
gravitational wave detectors such as LIGO and VIRGO are most sensitive.
In addition, efforts to discern between contributions from different
SASI modes -- e.g., sloshing and spiral -- may be affected, as well.

In light of the results presented here, further analysis is well motivated.
Our assumption of axisymmetry will need to be lifted, since the SASI is known
to have non-axisymmetric modes \citep{bm2007,bs2007,f2010,f2015}.
A full three-dimensional comparison of the SASI in NR and GR, such as the one
performed here for axisymmetry, should be conducted.
Further analyses may also include adding a third type of model that uses
NRHD and a GR monopole potential,
similar to what is done in several CCSN simulation
codes \citep[e.g.,][]{rj2002,ktf2018,sdb2019,bbh2020},
in order to discern whether the use
of an effective potential to capture the stronger gravitational fields present
in GR is able to better capture the SASI growth rate and subsequent evolution.
Of course, even if it does, nothing can replace a true GR implementation,
as we have done here, and as must be done in future CCSN models.

\centerline{\textbf{Acknowledgments}}\vspace{-1.0em}
\begin{acknowledgments}
S.J.D., E.E., and A.M. acknowledge support from the
National Science Foundation’s
Gravitational Physics Program under grant NSF PHY 1806692,
and E.E., A.M., and J.B.
acknowledge support from the National Science Foundation’s Gravitational
Physics Program under grant 2110177.
This research was supported in part by the Exascale Computing Project
(17-SC-20-SC), a collaborative effort of the U.S. Department of Energy Office
of Science and the National Nuclear Security Administration.
This work was conducted in part using
the resources of the Advanced Computing Center for Research and Education at
Vanderbilt University, Nashville, TN.
\end{acknowledgments}

\software
{
  \href{https://github.com/endeve/thornado}{thornado}
  \citep{pbe2021},
  \href{https://amrex-codes.github.io/}{AMReX}
  \citep{z2019},
  \href{https://matplotlib.org/}{Matplotlib}
  \citep{h2007},
  \href{http://www.numpy.org/}{NumPy}
  \citep{hmv2020},
  \href{http://www.scipy.org/}{SciPy}
  \citep{vgo2020},
  \href{https://yt-project.org/}{yt}
  \citep{tso2011}
}

\centerline{\textbf{Data Availability}}\vspace{0.5em}

The data underlying the figures and the code to reproduce the figures
are available on Zenodo under an open-source
Creative Commons Attribution license:
doi:\href{https://doi.org/10.5281/zenodo.10529594}{10.5281/zenodo.10529594}.
All simulations presented here have been performed
on the \texttt{MeshRefinement\_DG} branch of a
fork\footnote{https://github.com/dunhamsj/amrex} of version 21.09 of
\amrex{} \citep{amrex}.

\bibliography{bibfile}{}
\bibliographystyle{aasjournal}

\end{document}